\pdfoutput=1
\documentclass[iop]{emulateapj}
\slugcomment{}

\shorttitle{The study of a system of H\,{\sc ii} regions toward {\it l} = 24$\degr$.8, b = 0$\degr$.1}
\shortauthors{L.~K. Dewangan et al.}
\begin{document}

\title{The study of a system of H\,{\sc ii} regions toward {\it l} = 24$\degr$.8, b = 0$\degr$.1 
at the Galactic bar - Norma arm interface}
\author{L.~K. Dewangan\altaffilmark{1}, J.~S. Dhanya\altaffilmark{2}, D.~K. Ojha\altaffilmark{3}, and I. Zinchenko\altaffilmark{4}}
\email{lokeshd@prl.res.in}
\altaffiltext{1}{Physical Research Laboratory, Navrangpura, Ahmedabad - 380 009, India.}
\altaffiltext{2}{Malaviya National Institute of Technology (MNIT), Jaipur 302 017, India.}
\altaffiltext{3}{Department of Astronomy and Astrophysics, Tata Institute of Fundamental Research, Homi Bhabha Road, Mumbai 400 005, India.}
\altaffiltext{4}{Institute of Applied Physics of the Russian Academy of Sciences, 46 Ulyanov st., Nizhny Novgorod 603950, Russia.}
\begin{abstract}
To probe the star formation (SF) process, we present a thorough multi-wavelength investigation of several H\,{\sc ii} regions located toward {\it l} = 24$\degr$.8, b = 0$\degr$.1. 
A system of at least five H\,{\sc ii} regions including the mid-infrared bubble N36 (hereafter ``system N36"; extension $\sim$35 pc) is observationally investigated, and is located at a distance of 6.0 kpc. 
With this distance, the system N36 is found to be situated at the interface of the Galactic bar and the Norma Galactic arm in our Galaxy, 
where one may expect the collisions of molecular clouds due to the bar potential.
Each H\,{\sc ii} region (dynamical age $\sim$0.4--1.3 Myr) in the system is powered by an O-type star. The system contains 27 
ATLASGAL dust clumps at 870 $\mu$m. 
Several clumps are massive ($>$ 10$^{3}$ M$_{\odot}$), and have high bolometric luminosity ($>$ 10$^{3}$ L$_{\odot}$). Using the GRS $^{13}$CO line data, in the direction of the system N36, two velocity components are found around 109 and 113 km s$^{-1}$, and are linked in the velocity space. 
The morphological analysis of $^{13}$CO favours the presence of interacting molecular clouds in the system. 
Four H\,{\sc ii} regions and two 6.7 GHz masers are spatially observed at the common areas of the two clouds. 
The analysis of the {\it Spitzer} photometric data also traces the noticeable SF activity in the system. Considering the observational outcomes, the formation of O-type stars (including ongoing SF) in the system appears to be triggered by the collisions of molecular clouds at the bar-arm interface.
\end{abstract}
\keywords{dust, extinction -- HII regions -- ISM: clouds -- ISM: individual objects (N36) -- stars: formation -- stars: pre-main sequence} 
\section{Introduction}
\label{sec:intro}
The physical processes involved in the formation of massive O-type stars 
and their feedback mechanisms are still under debate \citep{zinnecker07,tan14}. 
The energetics of O-type stars can affect the origin of new low-mass and massive stars \citep{deharveng10}. 
The massive stars are often surrounded by the bubbles/rings/semi-ring-like structures traced at mid-infrared (MIR) 
8.0 $\mu$m \citep{churchwell06,churchwell07}, and are also associated 
with the extended radio continuum emission \citep[e.g.,][]{deharveng10}. 
Note that the majority of the studies related to the MIR bubbles are mainly carried out for a single H\,{\sc ii} region or 
several H\,{\sc ii} regions on scales of a few parsecs \citep[e.g.,][]{zinnecker07,deharveng10,rathborne11,tackenberg12,simpson12,kendrew12,thompson12,tan14,dewangan15a,dewangan15b,xu16}.
However, to our knowledge, in the Milky Way, there is still a limited detailed multi-wavelength study of large-scale systems ($>$ 25 pc) of several MIR bubbles/H\,{\sc ii} regions containing O-type stars, and hence the origin of such extended system of H\,{\sc ii} regions still remains unexplored. 
These systems could be candidates of ``mini-starburst" \citep[such as W43 `mini-starburst' region;][]{motte03}. 
With the availability of the radio recombination line (RRL) and continuum observations \citep[e.g.,][]{lockman89,condon98,anderson09,jones12}, the MIR survey \citep[e.g.,][]{benjamin03}, 
the dust continuum survey at 870 $\mu$m \citep[e.g.,][]{schuller09,urquhart18}, and the $^{13}$CO line survey \citep[e.g.,][]{jackson06,anderson09a}, it appears that the H\,{\sc ii} regions located toward the Galactic plane and the inner Galaxy are the promising sites to investigate the extended systems of O-type stars.
Such study will enable us to understand the physical conditions in a densely clustered environment linked with the luminous giant H\,{\sc ii} regions/massive star-forming complexes/mini-starburst candidates in the Galaxy. 
However, in particular, in the direction of the inner Galaxy, 
the investigation of an extended system of H\,{\sc ii} regions is often restricted by the near-far kinematic distance ambiguity \citep[e.g.,][]{anderson09,jones12,urquhart18}. 
In recent years, a significant effort has been devoted to resolve the distance ambiguity 
for H\,{\sc ii} regions in the inner Galaxy \citep[see][and references therein]{urquhart18}. 
In this work, we aim to observationally investigate a large-scale system/configuration of several H\,{\sc ii} regions powered by O-type stars and the origin of such a large system. 

Several extended H\,{\sc ii} regions are known in the direction of {\it l} = 23$\degr$.9 -- 24$\degr$.9 \citep[e.g.,][]{lockman89,kuchar97,kantharia07,jones12} 
(see Figure~\ref{ffg1}a using the VLA Galactic Plane Survey \citep[VGPS; $\lambda$ =21 cm;][]{stil06}).
Several MIR bubbles \citep[such as, N32, N33, N34, N35, and N36;][]{churchwell06,deharveng10,simpson12} are also reported in the selected 
longitude range, and are associated with the H\,{\sc ii} regions \citep[e.g.,][]{anderson09,urquhart13b}. 
In this longitude range, we have examined the spatial distribution of the 21 cm continuum emission and the dust continuum clumps at 870 $\mu$m \citep[from][]{urquhart18} (see Figure~\ref{ffg1}a). 
\citet{urquhart18} solved the distance ambiguity for the clumps using a combination of the H\,{\sc i} analysis, maser parallax, and spectroscopic measurements (see their paper for more details). 
Based on this exercise, three different ``clusters" of dense gas clumps are identified in the direction of 
{\it l} = 23$\degr$.9 -- 24$\degr$.9, which are located at distances of 5.8 kpc (radial velocity (V$_{lsr}$) range = [80, 100] km s$^{-1}$), 6.0 kpc (V$_{lsr}$ range = [104, 113] km s$^{-1}$), and $\sim$7.8 kpc (V$_{lsr}$ range = [115, 122] km s$^{-1}$).
One of the clusters is located toward the ionized regions hosting the MIR bubble N35 at a distance of $\sim$7.8 kpc, 
while the second one is found toward the H\,{\sc ii} regions associated with the MIR bubble N36 at a distance of 6 kpc.
The third one is seen in the direction of the bubbles N33 and N34 at a distance of 5.8 kpc. 
This analysis also indicates the existence of at least three different physical systems/velocity components in the projected sky area toward {\it l} = 23$\degr$.9 -- 24$\degr$.9. 
More recently, \citet{torii17x} studied the giant molecular cloud (GMC) associated with 
the MIR bubble N35, which is not physically associated with the bubbles N33, N34, and N36. 
Similarly, the bubble N36 does not have any physical connection with the bubbles N33 and N34.

In this paper, using multi-frequency data sets (see Table~\ref{xftab1}), we carry out a detailed investigation of the system of the H\,{\sc ii} regions associated with the MIR bubble N36 in the direction of {\it l} = 24$\degr$.6 -- 24$\degr$.9 (see an area enclosed by a box in Figure~\ref{ffg1}a). 
Hereafter, we refer this configuration to ``system N36", which is an extended 
and a single configuration of several dust clumps (having d = 6.0 kpc and V$_{lsr}$ range = [104, 113] km s$^{-1}$) and H\,{\sc ii} regions/21 cm bright continuum sources. 
Previously, the MIR bubble N36 \citep[{\it l} = 024$\degr$.837; {\it b} = +00$\degr$.090;][]{churchwell06} was characterized as a broken or incomplete ring with a mean radius and thickness of 2$\farcm$82 (or 4.9 pc at d = 6.0 kpc) and 0$\farcm$87 (or 1.5 pc at d = 6.0 kpc), respectively. In the direction of the bubble N36, \citet{kantharia07} reported an extended H\,{\sc ii} region G24.83+0.10 containing multiple radio peaks, and the H\,{\sc ii} region G24.83+0.10 was suggested to be powered by a single star of spectral type O5.5 \citep{kantharia07}. 
The bubble N36 was also suggested to be a good candidate for triggered massive star formation \citep[e.g.,][]{deharveng10}.

Using the Galactic structure models and different observational tracers (such as H\,{\sc ii} regions, GMCs, 6.7 GHz methanol masers etc.), several efforts have been made to understand the Galactic structure/spiral structure of our Galaxy \citep[e.g.][and references therein]{reid14,sato14,hou09,hou14,hu16}. 
Such studies help us to infer the physical association of the observed tracers with 
the spiral arms (e.g., Perseus, Carina-Sagittarius, Norma (and Outer arm), Crux-Scutum, Local) of the Milky Way \citep[e.g.][and references therein]{hou09,hou14,xu16b}. The locations of these spiral arms are presented in Figure~1 in \citet{elia17}.
\citet{sato14} listed the names of massive star-forming sites located in the Scutum Arm (see Table~2 in their paper). 
With the help of the literature, it appears that the system N36 ({\it l} = 24$\degr$.6 -- 24$\degr$.9 and d = 6.0 kpc) is not part of the Scutum Arm \citep[see Figure~5 in][]{hou14}, and is located in the inner Galaxy, where the Galactic bar meets the Norma Galactic arm in the Milky Way \citep[see also Figure~9 in][]{nguyen11}. 
It has also been suggested that the Norma Spiral arm harbors the most massive GMCs and the most luminous regions of massive star formation in the Milky Way \citep[e.g.,][]{bronfman08}. 
Using the {\it Spitzer} 3.6--8.0 $\mu$m data, \citet{benjamin05} reported the presence of a Galactic bar having a radius of 4.4 kpc in the Milky Way, 
which is oriented about 44$\degr$ relative to the Sun-Galactic center line. Together, the system N36 is an interesting target field, where one can carry out an investigation of the bar-arm interaction \citep[e.g.,][]{nguyen11,beuther17}. 

In this paper, we have focused our analysis in the system N36 to observationally understand 
the origin of the large-scale configuration of O-type stars. We also study the bar-arm interaction in our selected field. 
The study of dust clumps against the distribution of young stellar objects (YSOs) is also yet to be analyzed in the system N36.
Furthermore, we also examine the physical environment and the impact of massive stars in their vicinity. 
However, in the direction of {\it l} = 24$\degr$.6 -- 24$\degr$.9, {\it b} = $-$0$\degr$.2 -- 0$\degr$.2, a detailed distribution of molecular gas for the system N36 has not been investigated. 

The paper is organized in the following way. 
In Section~\ref{sec:obser}, we provide the information about the adopted data sets. 
Section~\ref{sec:data} presents the results related to the physical environment and point-like sources.  
In Section~\ref{sec:disc}, we discuss the possible star formation process operating in our selected target region. 
Finally, the main results are summarized in Section~\ref{sec:conc}.
%
\section{Data sets and analysis}
\label{sec:obser}
The multi-wavelength data sets adopted in this paper were retrieved from different surveys, which are listed in Table~\ref{xftab1} 
\citep[see][for more details]{dewangan17,dewangan17b,dewangan18}. 
 \begin{table*}
  \tiny
\setlength{\tabcolsep}{0.05in}
\centering
\caption{Multi-wavelength surveys adopted in this paper.}
\label{xftab1}
\begin{tabular}{lcccr}
\hline 
  Survey  &  Wavelength(s)       &  Resolution ($\arcsec$)        &  Reference \\   
\hline
\hline 
 NRAO VLA Sky Survey (NVSS)                                   & 21 cm                       & $\sim$46          & \citet{condon98}\\
 VLA Galactic Plane Survey (VGPS)                             & 21 cm                       & $\sim$60          & \citet{stil06}\\
 Galactic Ring Survey (GRS)                                                                   & 2.7 mm; $^{13}$CO (J = 1--0) & $\sim$45        &\citet{jackson06}\\
APEX Telescope Large Area Survey of the Galaxy (ATLASGAL)                 &870 $\mu$m                     & $\sim$19.2        &\citet{schuller09}\\
{\it Herschel} Infrared Galactic Plane Survey (Hi-GAL)                              &70, 160, 250, 350, 500 $\mu$m                     & $\sim$5.8, $\sim$12, $\sim$18, $\sim$25, $\sim$37         &\citet{molinari10}\\
{\it Spitzer} MIPS Inner Galactic Plane Survey (MIPSGAL)                                         &24 $\mu$m                     & $\sim$6         &\citet{carey05}\\ 
{\it Spitzer} Galactic Legacy Infrared Mid-Plane Survey Extraordinaire (GLIMPSE)       &3.6, 4.5, 5.8, 8.0  $\mu$m                   & $\sim$2, $\sim$2, $\sim$2, $\sim$2           &\citet{benjamin03}\\
SuperCOSMOS H$\alpha$ survey (SHS)                                                 &0.6563 $\mu$m                   &$\sim$1           &\citet{parker05}\\
\hline          
\end{tabular}
\end{table*}
\section{Results}
\label{sec:data}
\subsection{Physical environment of the system N36}
\label{subsec:radio}
\subsubsection{Multi-wavelength view}
\label{xxsubsec:radio}
\begin{figure*}
\epsscale{1}
\plotone{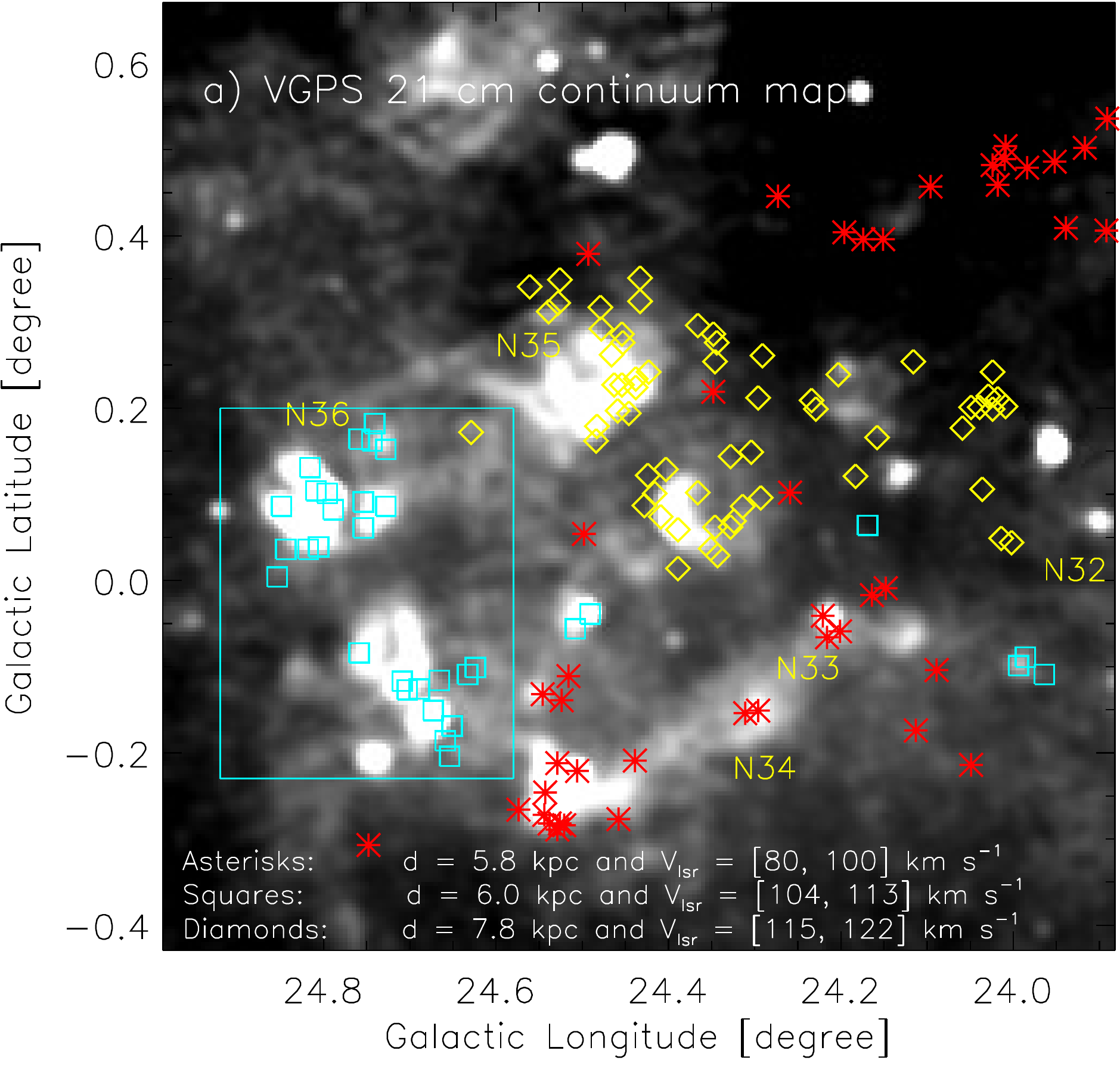}
\epsscale{1}
\plotone{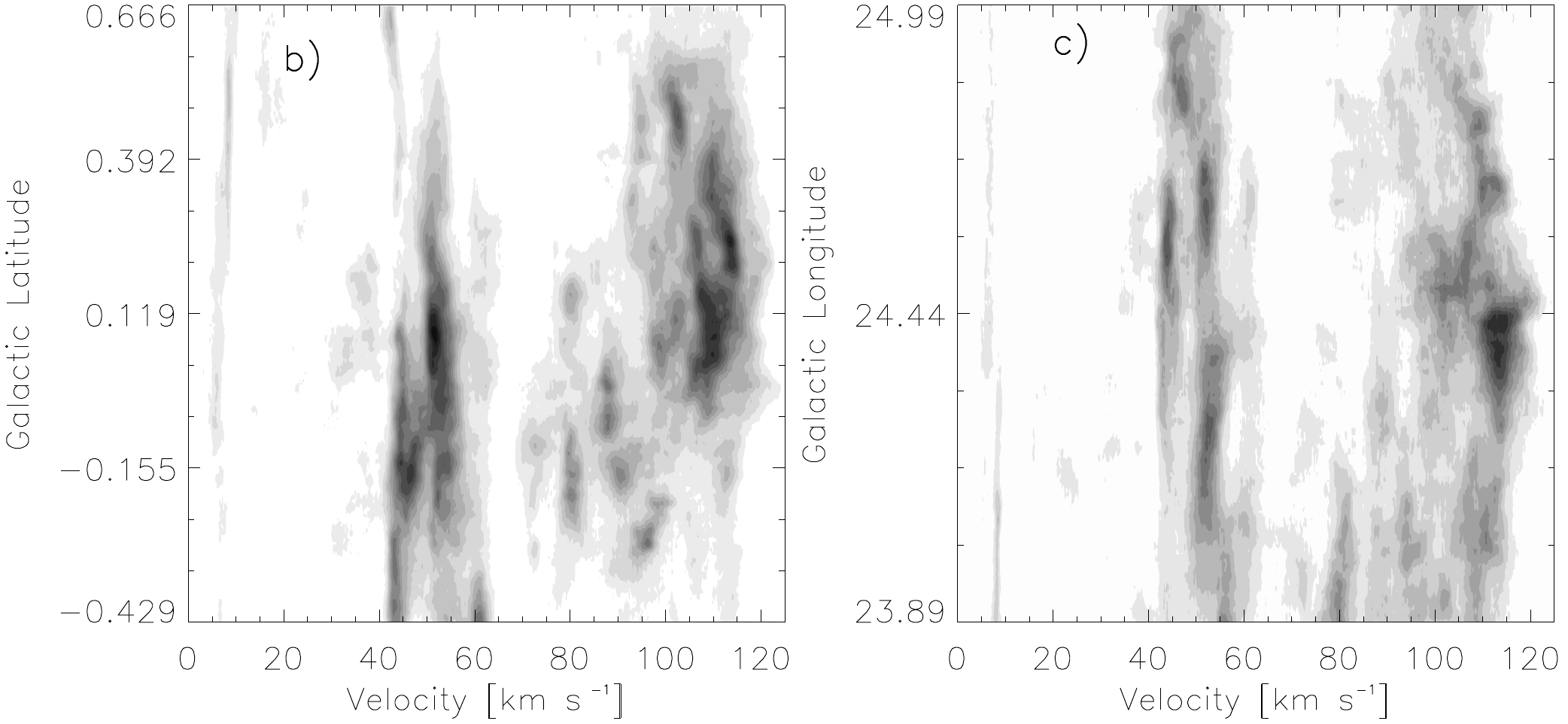}
\caption{a) VGPS 21-cm continuum emission map (selected area $\sim$1$\degr$.1 $\times$ 1$\degr$.1) toward {\it l} = 23$\degr$.9 -- 24$\degr$.9. 
The ATLASGAL dust continuum clumps at 870 $\mu$m \citep[from][]{urquhart18} are also overlaid on the map (see asterisk, square, and diamond symbols). 
The clumps highlighted with asterisks, squares, and diamonds are located at distances of 5.8, 6.0, and $\sim$7.8 kpc, respectively. 
Some previously known MIR bubbles \citep[such as N32, N33, N34, N35, and N36;][]{churchwell06} are also labeled in the map. 
The solid box (in cyan) refers to the area studied in this paper (see Figures~\ref{zfg1}a and~\ref{zfg1}b). b) Latitude-velocity map of $^{13}$CO. The $^{13}$CO emission is integrated over the longitude 
from 23$\degr$.89 to 24$\degr$.99 (see Figure~\ref{ffg1}a). c) Longitude-velocity map of $^{13}$CO. The $^{13}$CO emission is integrated over the latitude 
from $-$0$\degr$.429 to 0$\degr$.666 (see Figure~\ref{ffg1}a). The position-velocity maps are shown in the velocity range from 0 to 125 km s$^{-1}$. 
In all the panels, the longitude and latitude units are expressed in degrees.}
\label{ffg1}
\end{figure*}

Figure~\ref{ffg1}a displays an extended area containing three different ``clusters" of dense gas clumps in the direction of {\it l} = 23$\degr$.9 -- 24$\degr$.9 (see Section~\ref{sec:intro}). Based on the available observed parameters of the dust clumps (such as positions, distances, and radial velocities) in conjunction with the radio continuum data, we have investigated an extended system 
of H\,{\sc ii} regions in the direction of the inner Galaxy (i.e.``system N36"; extension $\sim$35 pc; see a box in Figure~\ref{ffg1}a). In Figures~\ref{ffg1}b and~\ref{ffg1}c, the position-velocity plots of the GRS $^{13}$CO (J=1--0) \citep[rms $\approx0.13$~K; velocity 
resolution of 0.21~km\,s$^{-1}$;][]{jackson06} trace different velocity components in the direction of a wide-field area around the system N36 (see Figure~\ref{ffg1}a). One can see the existence of complex velocity structures between 80 and 125 km s$^{-1}$ in the velocity space. 
These plots are displayed in a velocity range from 0 to 125 km s$^{-1}$. 
Note that the molecular gas in the direction of the system N36 is examined in a velocity range of 
[102, 116] km s$^{-1}$.

\begin{figure*}
\epsscale{1.15}
\plotone{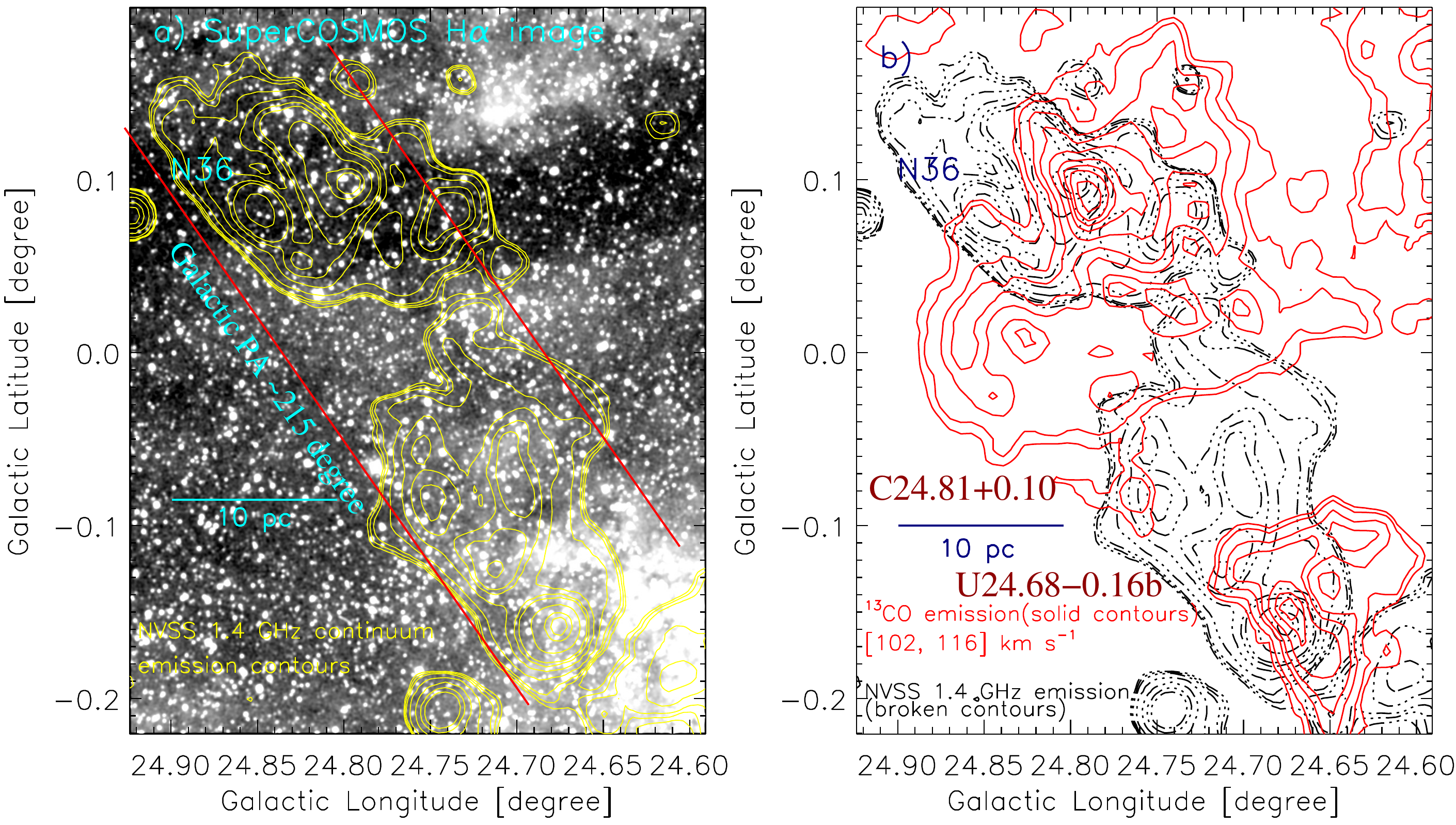}
\caption{a) Overlay of the NVSS 1.4 GHz continuum emission contours (1$\sigma$ $\sim$0.45 mJy/beam) on the SHS H$\alpha$ image. The contours (in yellow) are shown with the 
levels of 2.5 (5.5$\sigma$), 3.15, 4, 7.5, 18, 38, 60, 180, 450, and 650 mJy/beam. Each solid line (in red) is highlighted with a GPA of 215$\degr$. The SHS H$\alpha$ image is processed through a Gaussian smoothing function with a width of 4 pixels. 
b) Distribution of the $^{13}$CO(J =1$-$0) emission (in red) at [102, 116] km s$^{-1}$ against the NVSS 1.4 GHz continuum emission (in black). The $^{13}$CO emission contour levels (in red) are drawn with the levels of 12, 15, 22, 28, 38, 50, 58, 65, and 75 K km s$^{-1}$. The NVSS broken 
contours (in black) are the same as in Figure~\ref{zfg1}a. The scale bar referring to 10 pc (at a distance of 6.0 kpc) is shown in the both the panels.}
\label{zfg1}
\end{figure*}

In Figures~\ref{zfg1}a and~\ref{zfg1}b, we present a zoomed-in view of the system N36 using 
the SHS H$\alpha$, GRS $^{13}$CO, and NVSS 1.4 GHz images 
(size $\sim$0$\degr$.33 (34.6 pc) $\times$ 0$\degr$.42 (44.0 pc)). 
Figure~\ref{zfg1}a displays the SHS H$\alpha$ image overlaid with the NVSS 1.4 GHz continuum emission contours. 
The NVSS map reveals extended ionized emissions in the system N36, which appear parallel to a line having a Galactic position angle (GPA) of 215$\degr$ 
(see solid red lines in Figure~\ref{zfg1}a). 
There is no extended diffuse emission seen in the H$\alpha$ image. However, in the Galactic southern side, some noticeable H$\alpha$ emission is observed. 
It appears that dust extinction is affecting the optical emission from N36. Figure~\ref{zfg1}b shows the $^{13}$CO emission contours against the NVSS 1.4 GHz radio emission. 
The integrated intensity map of $^{13}$CO from 102 to 116 km s$^{-1}$ traces a spatial distribution of the molecular gas toward the system N36. 
In the literature, the clouds C24.81+0.10 (V$_{lsr}$ = 108.3 km s$^{-1}$; $\Delta$V = 8.2 km s$^{-1}$) and U24.68$-$0.16b (V$_{lsr}$ = 112.5 km s$^{-1}$; $\Delta$V = 5.1 km s$^{-1}$) 
are reported toward our selected field. These clouds are also labeled in Figure~\ref{zfg1}b. In the system N36, the cloud C24.81+0.10 is observed in the Galactic northern direction, whereas the cloud U24.68$-$0.16b is found in the Galactic southern side. A detailed study of the molecular gas is presented in Section~\ref{sec:coem}. 

\begin{figure*}
\epsscale{1.15}
\plotone{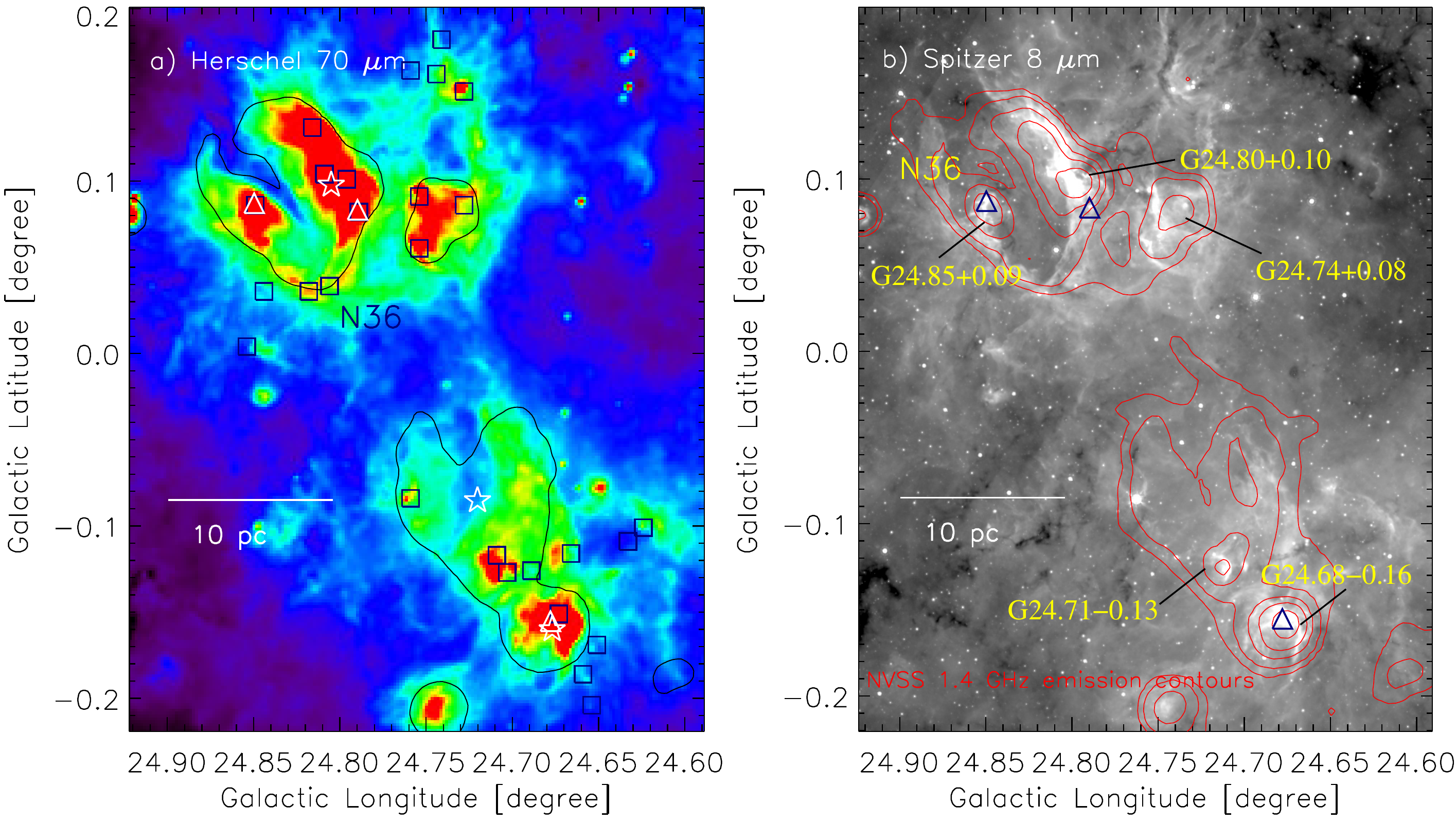}
\caption{Distribution of warm dust, cold dust, and ionized emission toward the selected region 
in this paper (size $\sim$0$\degr$.33 $\times$ 0$\degr$.42 ($\sim$34.6 pc $\times$ 44.0 pc 
at a distance of 6.0 kpc); central coordinates: $l$ = 24$\degr$.757; $b$ = $-$0$\degr$.011). 
a) Overlay of the ATLASGAL dust continuum clumps at 870 $\mu$m on a false color {\it Herschel} 70 $\mu$m image. 
The NVSS 1.4 GHz contour is also shown with a level of 16 mJy/beam. 
Twenty seven ATLASGAL clumps highlighted with squares (V$_{lsr}$ range $\sim$105--114 km s$^{-1}$) 
are found at a distance of 6.0 kpc (see Table~\ref{tab2}). 
Three stars (in white) indicate the positions of the RRL observations \citep[V$_{lsr}$ range $\sim$108--112 km s$^{-1}$;][]{lockman89}. 
b) Overlay of the NVSS 1.4 GHz emission contours (in red) on the {\it Spitzer} 8.0 $\mu$m image. 
The NVSS 1.4 GHz continuum contours are shown with the levels of 7.4, 18.6, 74.2, 185.6, and 464.1 mJy/beam. Several ionized regions (i.e G24.83+0.10 (hosting G24.85+0.09 and G24.80+0.10), G24.74+0.08, 
G24.71-0.13, and G24.68-0.16) are also marked in the figure. 
In both the panels, the positions of the Class~II 6.7 GHz methanol masers \citep[from][]{szymczak12} are shown by triangles (V$_{lsr}$ range $\sim$110--116 km s$^{-1}$).}
\label{sg1}
\end{figure*}

Figures~\ref{sg1}a and~\ref{sg1}b show a spatial view of the system N36 using the {\it Herschel} 70 $\mu$m and {\it Spitzer} 8.0 $\mu$m images, respectively. The images at 8.0 and 70 $\mu$m are overlaid with the NVSS radio continuum emission. 
The 70 $\mu$m image is also overlaid with the ATLASGAL dust continuum clumps at 870 $\mu$m. 
Table~\ref{tab2} lists the physical parameters of the ATLASGAL clumps
(i.e. peak flux density, integrated flux density, radial velocity, distance, effective radius, dust temperature, bolometric luminosity, clump mass, and molecular hydrogen column density), 
which are obtained from \citet{urquhart18} (see their paper for more details). 
Using the molecular line observations (e.g., CO, NH$_{3}$, CS etc), \citet{urquhart18} reported the measured 
radial velocity ranges of the ATLASGAL clumps to be 104--113 km s$^{-1}$ (see Table~\ref{tab2}). One can find that these 27 clumps (i.e. c1--c27; see squares in Figure~\ref{sg1}a) are located at a distance of 6.0 kpc (see Table~\ref{tab2}). Three positions of the RRL observations \citep[from][]{lockman89} are also marked in Figure~\ref{sg1}a (see star symbols in Figure~\ref{sg1}a). These ionized regions are well depicted in the ionized gas velocity range of 108--112 km s$^{-1}$ \citep[e.g.,][]{lockman89,anderson09}.
Several H\,{\sc ii} regions (such as G24.85+0.09, G24.80+0.10, G24.74+0.08, G24.71-0.13, and G24.68-0.16) are also highlighted in Figure~\ref{sg1}b 
\citep[e.g.,][]{kantharia07}. The two compact/ultra-compact H\,{\sc ii} regions G24.85+0.09 and G24.80+0.10 are seen with an extended H\,{\sc ii} region G24.83+0.10. All these H\,{\sc ii} regions are members of the system N36, and are physically associated with each other.
We have also marked the observed positions of the Class~II 6.7 GHz methanol masers \citep[from][]{szymczak12} in Figures~\ref{sg1}a and~\ref{sg1}b, which are well depicted in a velocity range of 110--115 km s$^{-1}$. These masers are also physically associated with the radio continuum sources (such as G24.85+0.09, G24.80+0.10, and G24.68-0.16), further indicating the presence of massive stars. 
It is supported by the fact that the 6.7 GHz methanol masers are a reliable tracer of massive YSOs \citep[e.g.,][]{walsh98,urquhart13a}. 

\begin{figure*}
\epsscale{1.15}
\plotone{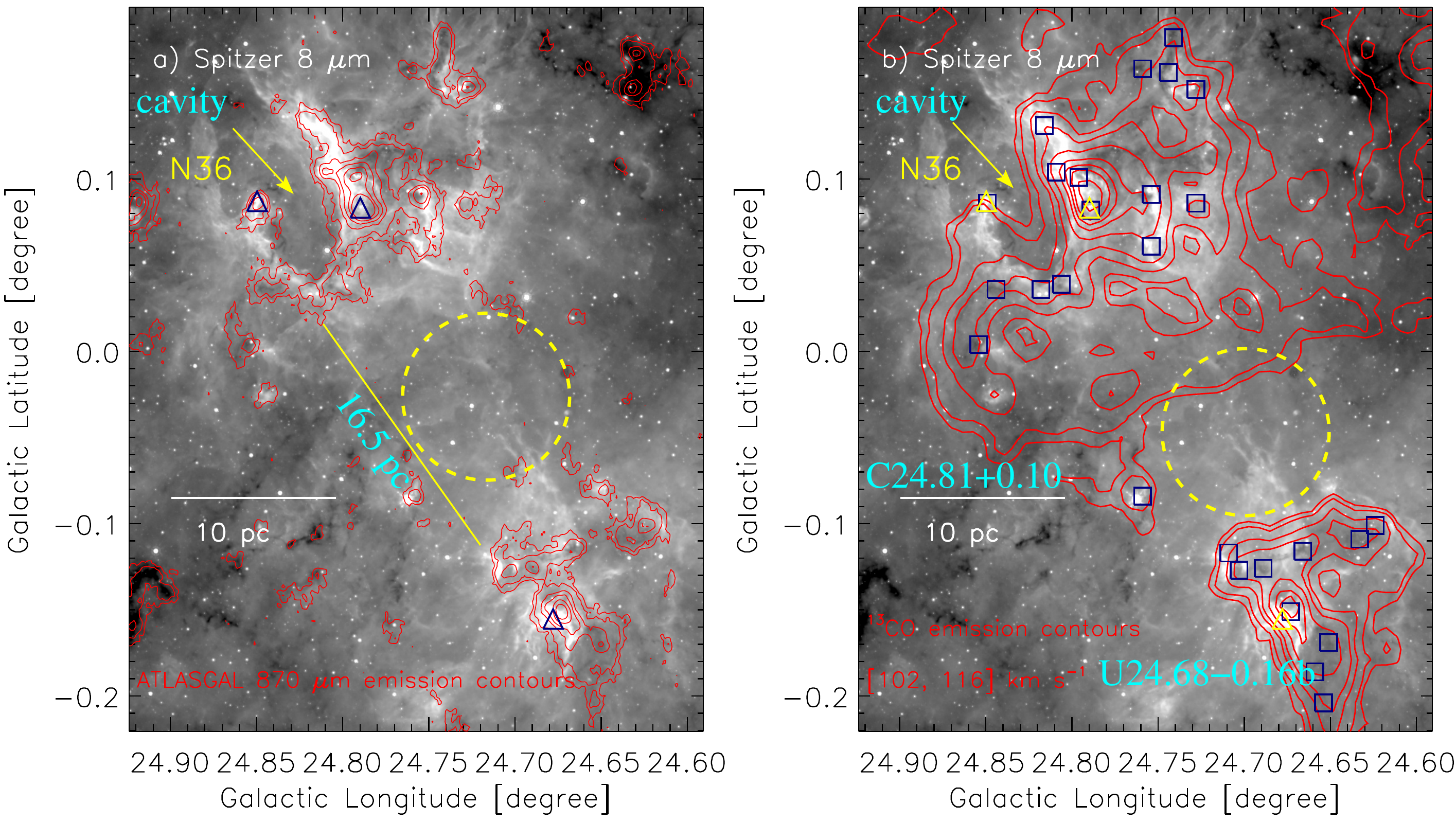}
\caption{a) Overlay of the ATLASGAL 870 $\mu$m continuum emission on the {\it Spitzer} 8.0 $\mu$m image. 
The contour levels of the ATLASGAL 870 $\mu$m emission are 15.2 Jy/beam $\times$ (0.01, 0.02, 0.05, 0.08, 0.13, and 0.2). 
b) Overlay of the $^{13}$CO(J =1$-$0) emission at [102, 116] km s$^{-1}$ and the ATLASGAL 870 $\mu$m continuum clumps 
on the {\it Spitzer} 8.0 $\mu$m image. 
The contour levels of the $^{13}$CO emission are 12, 15, 22, 28, 38, 50, 58, 65, and 75 K km s$^{-1}$ (see also Figure~\ref{zfg1}b). 
Other symbols are the same as in Figure~\ref{sg1}. 
In each panel, a dashed big circle (in yellow) indicates an area without any molecular emission.}
\label{sg2}
\end{figure*}

In Figures~\ref{sg2}a and~\ref{sg2}b, we have superimposed the ATLASGAL 870 $\mu$m continuum contours and the $^{13}$CO emission contours on the {\it Spitzer} 8.0 $\mu$m image, respectively. 
The integrated $^{13}$CO emission contours are shown here only for the comparison purpose. Two previously known clouds (i.e., C24.81+0.10 and U24.68$-$0.16b) are also labeled in Figure~\ref{sg2}b. One of the clouds (i.e. C24.81+0.10) is 
associated with the H\,{\sc ii} regions G24.83+0.10 (containing G24.85+0.09 and G24.80+0.10) and G24.74+0.08, while the other one (i.e. U24.68$-$0.16b) 
contains the H\,{\sc ii} regions G24.71-0.13, and G24.68-0.16. 
In Figure~\ref{sg2}b, the positions of the ATLASGAL clumps are also marked, allowing to infer the dense regions in the molecular clouds. 
An extended dust continuum emission at 870 $\mu$m is detected in the direction of each cloud (see Figures~\ref{sg2}a and~\ref{sg2}b). 
One can also note that the H\,{\sc ii} region G24.71-0.13 is seen between the clouds C24.81+0.10 and U24.68$-$0.16b (see also Figure~\ref{zfg1}b), where no dense materials at 870 $\mu$m and molecular emission are found (see a dashed big circle in Figures~\ref{sg2}a and~\ref{sg2}b). 
The dust continuum emission, 6.7 GHz methanol masers, and molecular emission are prominently observed at the edges of the bubble N36. In the direction of the bubble N36, a cavity-like feature is also highlighted in Figures~\ref{sg2}a and~\ref{sg2}b, where the dust continuum emission and the molecular gas are not observed. Basically, the cavity-like feature represents an interior of 
the MIR bubble N36 (having an average radius of 4.9 pc).  
The image at 70 $\mu$m depicts the warm dust emission, whereas the 8.0 $\mu$m band contains the 7.7 and 8.6 $\mu$m polycyclic aromatic hydrocarbon (PAH) emission (including the continuum). 
Bright and extended emission seen in the 8.0 and 70 $\mu$m continuum images is in good agreement, and is found at the locations of the ionized regions. Hence, concerning the system N36, the morphological agreement is seen in all the multi-wavelength images. 
The spatial distribution of the molecular gas, dust emission, and ionized gas indicates the impact of the H\,{\sc ii} regions in the system N36. 
\subsubsection{Temperature map and powering sources}
\begin{figure*}
\epsscale{1.15}
\plotone{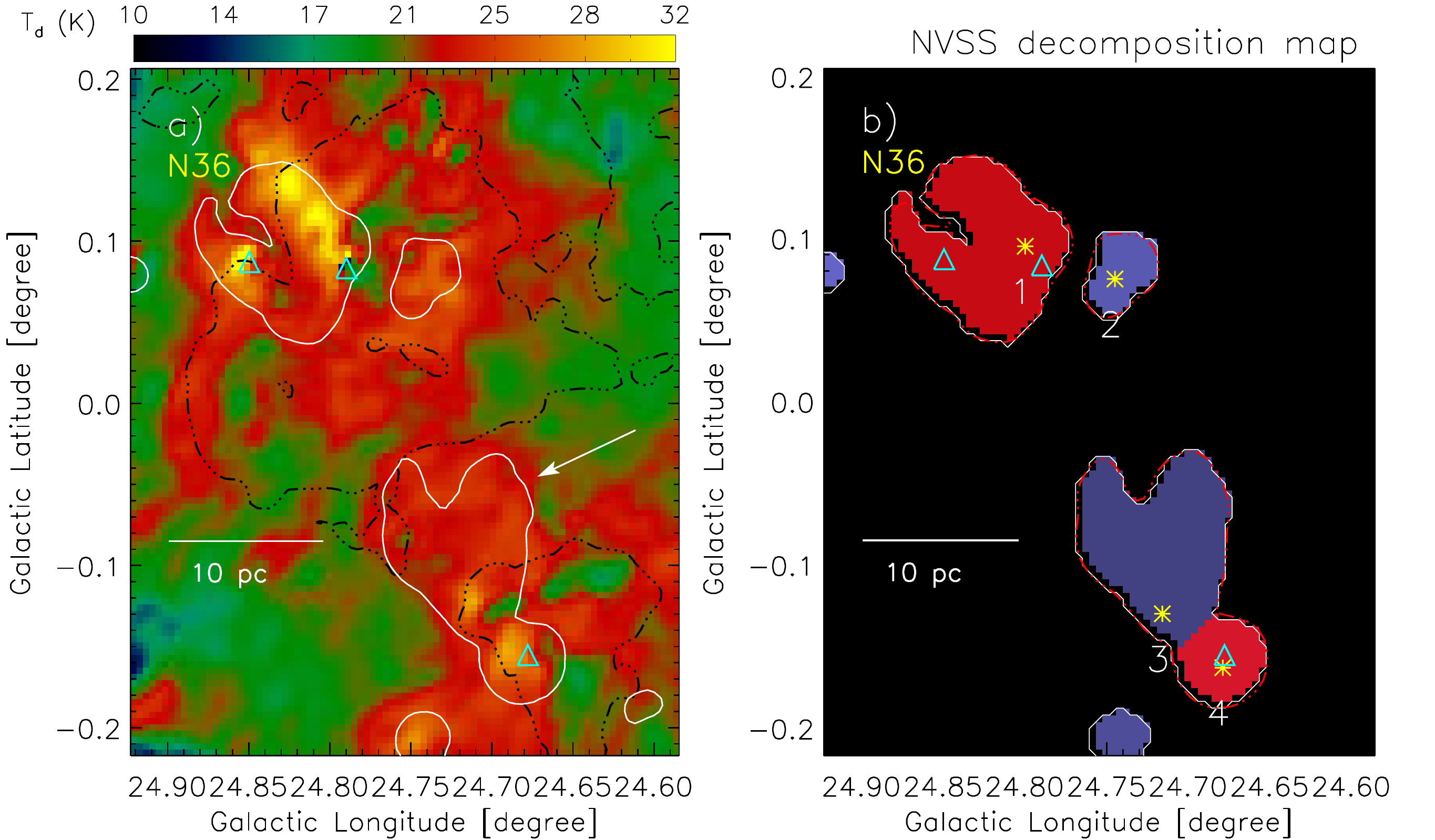}
\caption{a) Overlay of the NVSS 1.4 GHz emission contour (in white) on the {\it Herschel} temperature map (resolution $\sim$37$''$). 
The $^{13}$CO (J = 1-0) emission contour at [102, 116] km s$^{-1}$ (in black) is also shown with the level of 12 K km s$^{-1}$.
An arrow indicates the presence of ionized emission between molecular clouds. 
b) Clumpfind decomposition of the NVSS continuum emission. The boundary of each identified H\,{\sc ii} region in the NVSS 1.4 GHz map is 
highlighted along with its corresponding ID and position (see asterisks and also Table~\ref{tab3}). 
The NVSS 1.4 GHz emission contour with the level of 0.016 Jy/beam is also shown in each panel. 
In both the panels, the positions of the Class~II 6.7 GHz methanol masers are shown by triangles (in cyan).}
\label{sg3}
\end{figure*}

The {\it Herschel} temperature map of the system N36 is shown in Figure~\ref{sg3}a.
Following the methods given in \citet{mallick15}, the map is produced using the {\it Herschel} 160--500 $\mu$m data. 
The temperature map is overlaid with the NVSS radio continuum emission, indicating that the majority of the H\,{\sc ii} regions are 
associated with the warmer emission ($T_\mathrm{d}$ $\sim$23-32 K). 
There is also an extended temperature structure seen in the system N36.
The integrated $^{13}$CO emission at [102, 116] km s$^{-1}$ is also superimposed on the temperature map. 

Based on the NVSS radio map (1$\sigma$ $\sim$0.45 mJy/beam), in the system N36, we have selected four H\,{\sc ii} regions (labeled as 1-4) and their positions are marked in Figure~\ref{sg3}b. 
The IDs 1, 2, 3, and 4 refer to the H\,{\sc ii} regions G24.83+0.10 (containing G24.85+0.09 and G24.80+0.10), 
G24.74+0.08, G24.71-0.13, and G24.68-0.16, respectively. The boundary of each H\,{\sc ii} region is highlighted in the 1.4 GHz map (see Figure~\ref{sg3}b), which is an outcome of the ``{\it clumpfind}" IDL program \citep{williams94}. 
The 6.7 GHz methanol masers are seen in the H\,{\sc ii} regions 1 (i.e. G24.85+0.09 and G24.80+0.10) and 4 (i.e. G24.68-0.16). 
Following the analysis given in \citet[][and see references therein]{dewangan17}, 
the values of Lyman continuum photon ($N_\mathrm{uv}$) and dynamical age ($t_\mathrm{dyn}$) are estimated for each H\,{\sc ii} region using the NVSS radio continuum data, and are listed in Table~\ref{tab3}. 
This exercise reveals the spectral types of the ionizing stars of 
the H\,{\sc ii} regions to be O type stars, which are in almost agreement with the results of \citet{kantharia07}. The $t_\mathrm{dyn}$ values of these H\,{\sc ii} regions vary between 0.4 and 1.3 Myr (for an initial gas density n$_{0}$ = 10$^{3}$ cm$^{-3}$). 

Using the NVSS 1.4 GHz map, we have also estimated the physical properties of the ionized gas (i.e., electron density (n$_{e}$), emission measure (EM), and mass of ionized hydrogen (M$_{HII}$)). 
These parameters are also listed in Table~\ref{tab3}, and are derived based on the equations given in \citet{panagia78} for a roughly spherical geometry case. 
All the H\,{\sc ii} regions have high EM values ($>$ 10$^{4}$ cm$^{-6}$ pc). 
The H\,{\sc ii} regions 1 and 3 have higher values of M$_{HII}$ compared to other two H\,{\sc ii} regions (i.e., IDs 2 and 4; see Table~\ref{tab3}). 
\subsubsection{Distribution of clump parameters}
\begin{figure*}
\epsscale{1.1}
\plotone{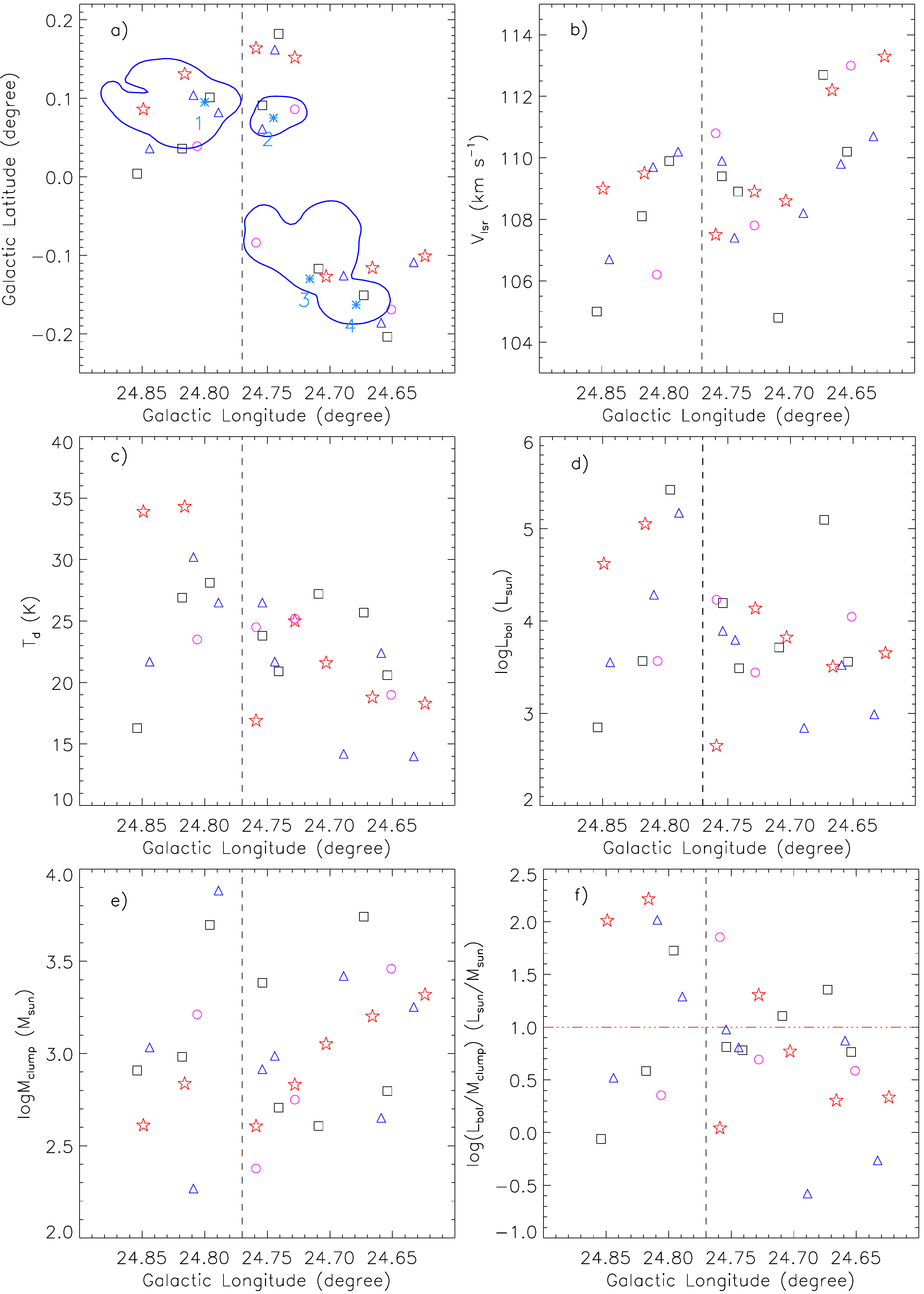}
\caption{a) Distribution of 27 ATLASGAL dust continuum clumps at 870 $\mu$m toward the selected region in this paper (see squares in Figure~\ref{sg1}a). 
The positions of four H\,{\sc ii} regions are highlighted by asterisks and labeled in the figure (see Figure~\ref{sg3}b). 
The NVSS emission toward these ionized regions is shown by blue curves. 
b-c-d-e-f) Distribution of the radial velocity, dust temperature, bolometric luminosity, mass, and ratio of the bolometric luminosity of clumps against the Galactic longitude. 
In each panel, four different symbols (i.e. squares, circles, stars, and triangles) indicate the dust clumps.}
\label{sg4}
\end{figure*}

We have identified 27 ATLASGAL dust continuum clumps at 870 $\mu$m \citep[from][]{urquhart18} toward the system N36 (see Figure~\ref{sg4}a and also Table~\ref{tab2}). 
We find that the dust clumps c1-c10 and c18 are embedded in the cloud U24.68$-$0.16b, while the other ATLASGAL clumps (c11-c17 and c19-c27) are seen toward the cloud C24.81+0.10 (see Figure~\ref{sg2}b). It implies that the clouds contain several massive clumps (see Table~\ref{tab2}). 
One can also notice that the majority of the clumps are distributed toward the four ionized regions (see Figure~\ref{sg4}a). However, some clumps are also seen away from the NVSS radio continuum emission (see Figure~\ref{sg4}a). 
Figure~\ref{sg4}b shows the distribution of the radial velocity of clumps against the Galactic longitude. 
The radial velocity of clumps toward the ionized regions 1, 2, and 3 are found between 104 and 111 km s$^{-1}$, 
while the clumps seen toward the ionized region 4 are depicted with 
the radial velocity of 109--114 km s$^{-1}$. These results indicate the presence of two velocity components in the system N36.
Figure~\ref{sg4}c presents the distribution of the dust temperature of clumps 
against the Galactic longitude. The figure shows a dust temperature range of $\sim$23--35 K toward the clumps, 
which are located toward the H\,{\sc ii} regions. This result is in agreement with the outcome 
of the {\it Herschel} temperature map (see Figure~\ref{sg3}a).  In Figure~\ref{sg4}d, we show the distribution of the bolometric luminosity of clumps against the Galactic longitude, which reveals the presence of high luminosity clumps ($>$ 10$^{3}$ L$_{\odot}$) toward the ionized regions in the system N36. In Figure~\ref{sg4}e, we display the distribution of clump masses against the Galactic longitude. The clump masses vary between 185 and 7635 M$_{\odot}$. Massive clumps ($>$ 10$^{3}$ M$_{\odot}$) are mainly seen toward all the ionized regions. 
Figure~\ref{sg4}f shows the distribution of the ratio of the bolometric luminosity and mass of clumps (i.e., L$_{bol}$/M$_{clump}$) against the Galactic longitude. The ratio L$_{bol}$/M$_{clump}$ of clumps is considered as an indicator of clumps evolution \citep[e.g.,][]{molinari16}. The clumps with L$_{bol}$/M$_{clump}$ $>$ 10 indicate their association with the H\,{\sc ii} regions \citep[e.g.,][]{molinari16}.
We find 9 ATLASGAL clumps having L$_{bol}$/M$_{clump}$ $>$ 10 in the system N36 (see Figure~\ref{sg4}f). 
The hydrogen column density range varies between 8.0 $\times$ 10$^{21}$ and 1.0 $\times$ 10$^{23}$ cm$^{-2}$. 
Furthermore, the 6.7 GHz methanol masers are associated with the 
dust clumps c7, c19, and c26, which have high bolometric luminosities (i.e., 4--15 $\times$ 10$^{4}$ L$_{\odot}$) 
and clump masses (i.e., 400--7635 M$_{\odot}$). 
\subsection{Kinematics of molecular gas}
\label{sec:coem} 
In this section, to probe the kinematics of molecular gas in the system N36, the study of molecular gas is carried out 
using the GRS $^{13}$CO (J=1--0) line data.  

\begin{figure*}
\epsscale{1}
\plotone{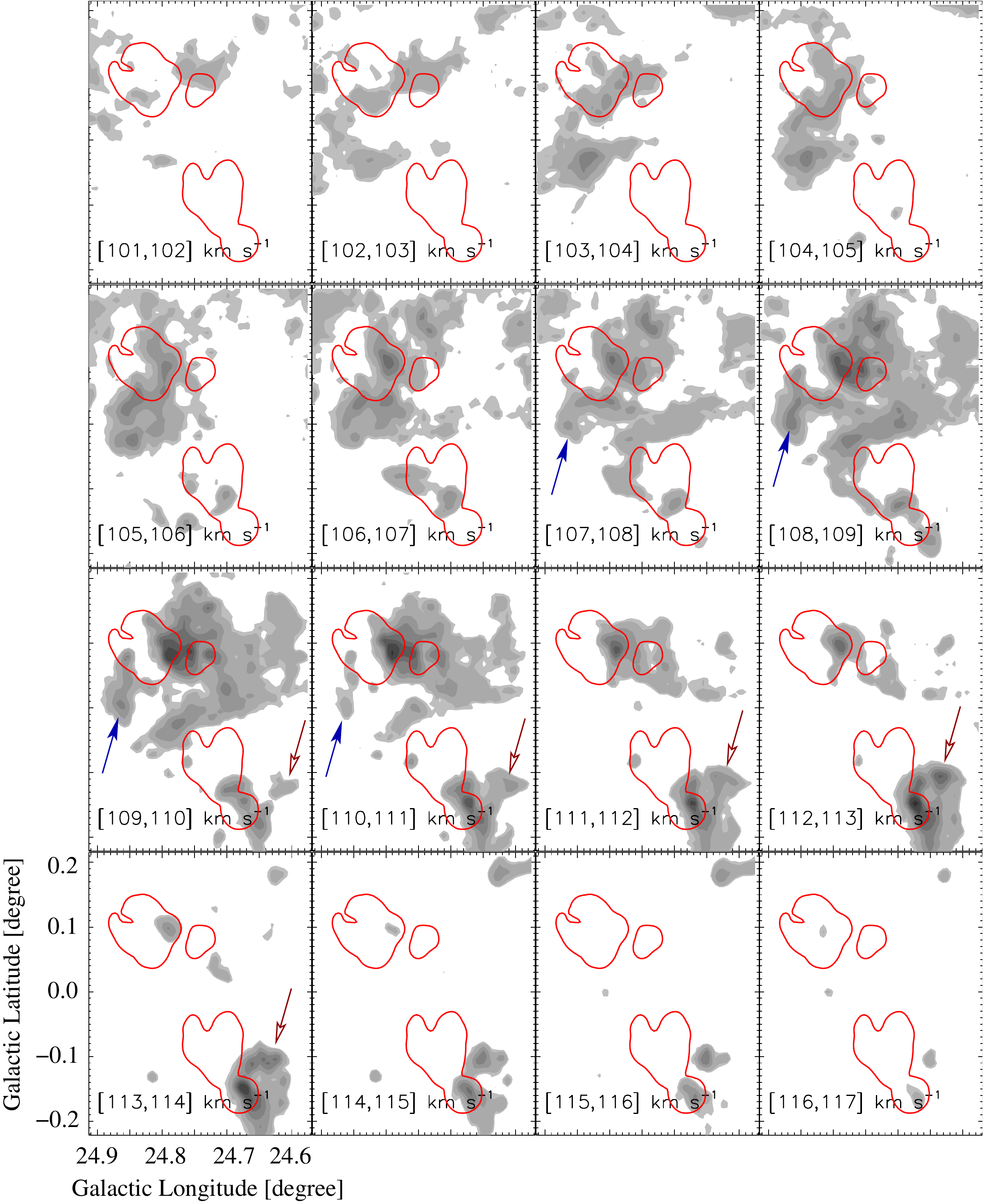}
\caption{Velocity channel maps of the $^{13}$CO(J =1$-$0) emission.
The molecular emission is integrated over a velocity interval, which is given in each panel (in km s$^{-1}$). 
The contour levels are 1.5, 2, 4, 6, 8, 10, 12, and 15 K km s$^{-1}$. 
In each panel, the red curves are the same as in Figure~\ref{sg4}a. Molecular gas in two clouds is indicated by arrows.}
\label{sg5}
\end{figure*}

\begin{figure*}
\epsscale{1.15}
\plotone{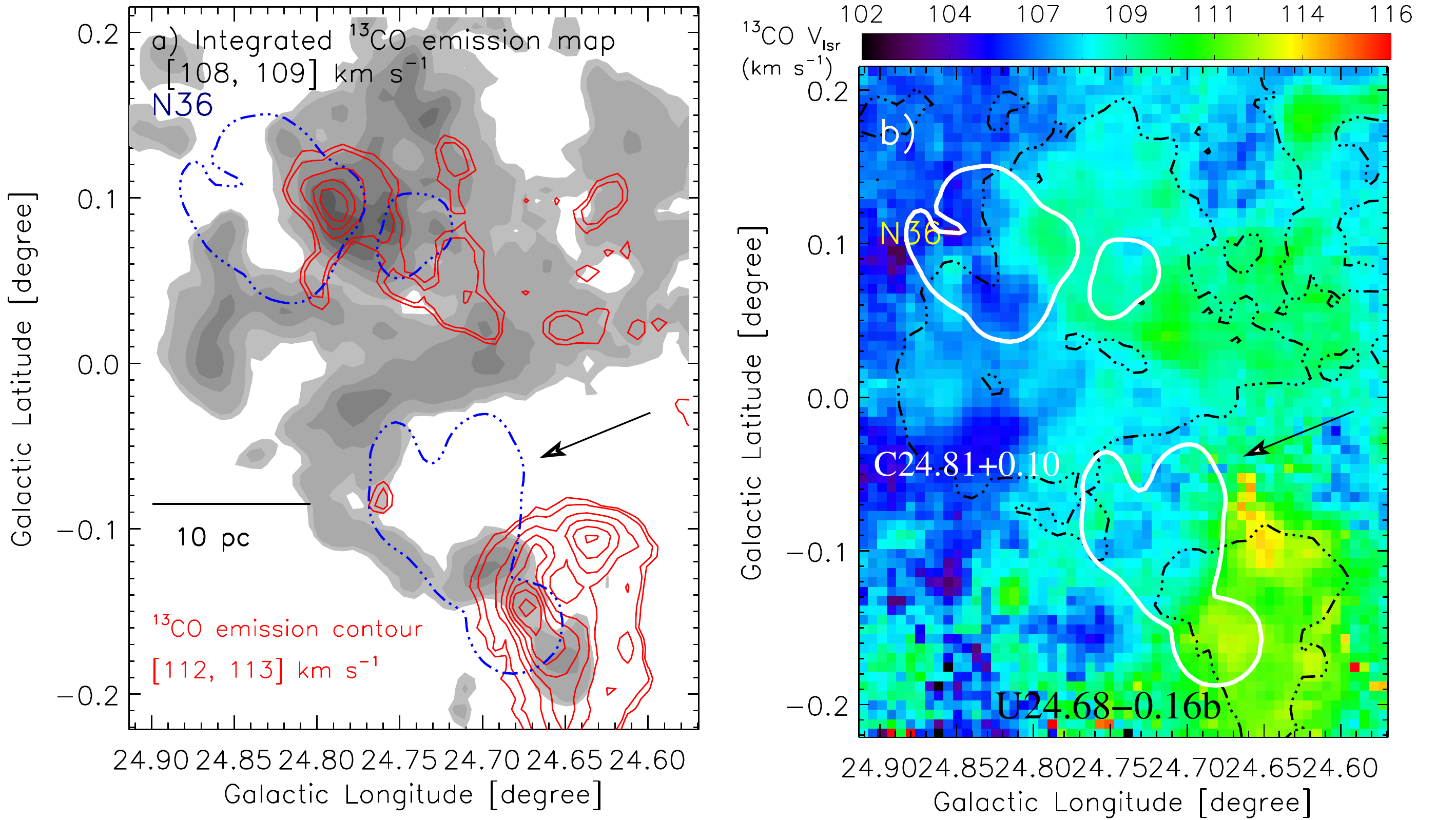}
\caption{a) Spatial distribution of the $^{13}$CO emission integrated over two different velocity ranges (i.e. 108--109 and 112--113 km s$^{-1}$; see also Figure~\ref{sg5}).
The contour levels of the background molecular map (at 108--109 km s$^{-1}$) and molecular contours (in red; at 112--113 km s$^{-1}$) are the same as shown in Figure~\ref{sg5}. 
b) Overlay of the $^{13}$CO emission at [102, 116] km s$^{-1}$ on the GRS $^{13}$CO first-order moment map. 
The bar indicates the mean $V_\mathrm{lsr}$ in km s$^{-1}$. 
The $^{13}$CO emission contour is overlaid with a level of 12 K km s$^{-1}$.
In both the panels, the NVSS 1.4 GHz emission is highlighted by curves, which are the same as in Figure~\ref{sg4}a. In each panel, an arrow indicates the presence of an extended ionized emission between molecular clouds.}
\label{sg6}
\end{figure*}

In the direction of the system N36, the radial velocity (V$_{lsr}$) ranges of the ionized gas, 
the molecular gas, and the 6.7 GHz methanol masers are reported to be 108--112, 105--114, and 110--116 km s$^{-1}$, respectively. 
Figure~\ref{sg5} shows the integrated $^{13}$CO velocity channel maps (starting from 101 km s$^{-1}$ at intervals of 1 km s$^{-1}$), where the locations of the ionized emission are also marked. 
The velocity channel maps trace 
at least two molecular components (around 109 and 113 km s$^{-1}$) along the line of sight in the system N36 (see arrows in Figure~\ref{sg5}). 
In the system N36, there is an extended GMC seen in the velocity between 103 and 110 km s$^{-1}$, and there exists also another cloud component in the velocity between 111 and 115 km s$^{-1}$. In the support of these results, in Figure~\ref{sg6}a, we have presented integrated velocity maps at [108, 109] km s$^{-1}$ and [112, 113] km s$^{-1}$ 
overlaid with the radio continuum emission. 
There are some areas in the system N36, where the molecular emissions observed in both the integrated velocity maps are distributed. 
In Figure~\ref{sg6}b, we show the $^{13}$CO first-order moment map, revealing the mean $V_\mathrm{lsr}$ of $^{13}$CO at each grid point. 
The map is also superimposed with the integrated molecular emission contour at [102, 116] km s$^{-1}$, as shown in Figures~\ref{zfg1}b and~\ref{sg2}b. 
The velocity gradients are observed in the clouds C24.81+0.10 and U24.68$-$0.16b.

\begin{figure*}
\epsscale{1}
\plotone{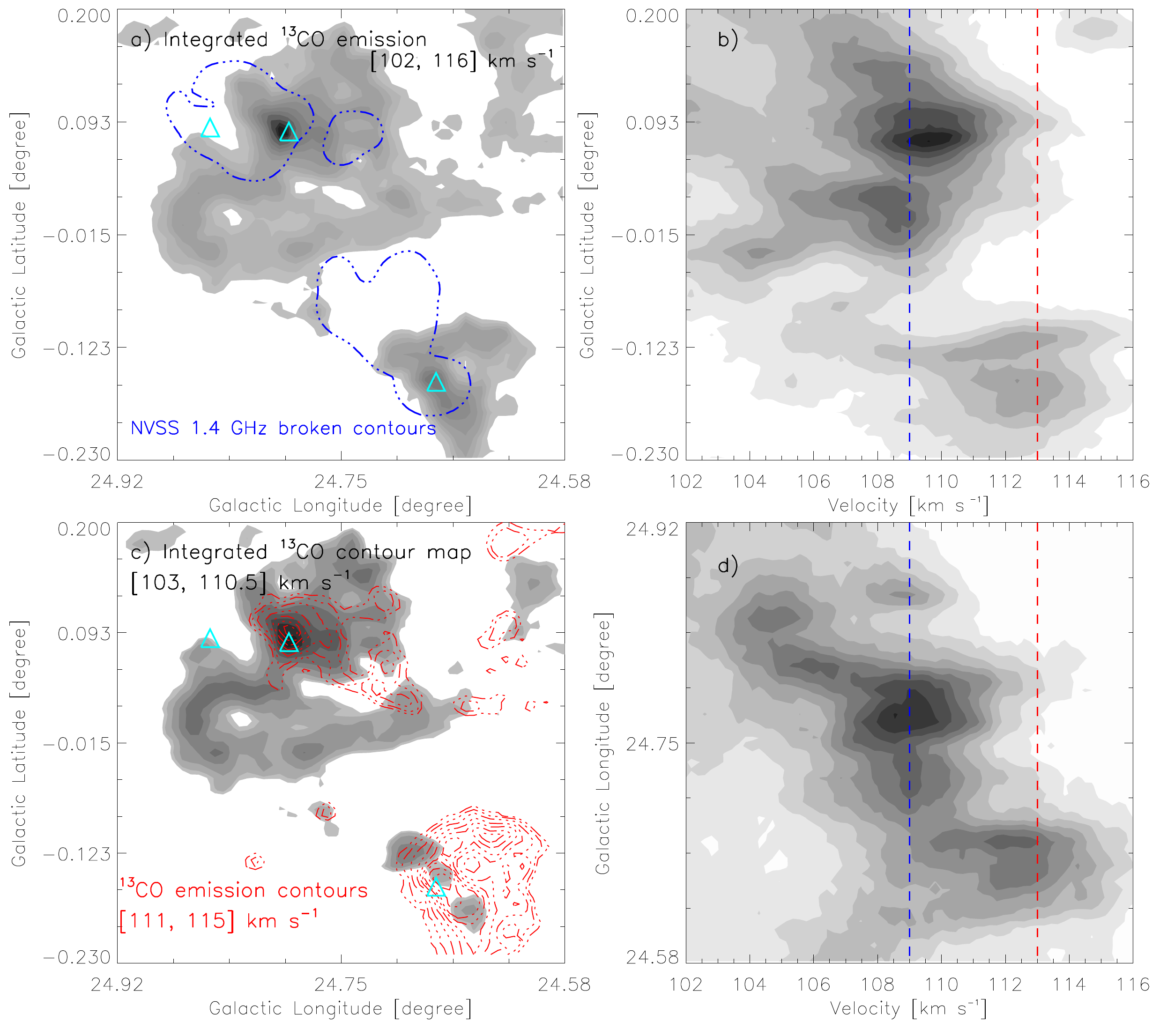}
\caption{a) Integrated intensity map of $^{13}$CO (J = 1-0) from 102 to 116 km s$^{-1}$. 
The$^{13}$CO emission contour levels are the same as in Figure~\ref{sg2}b. 
The broken blue curves are the same as in Figure~\ref{sg4}a. 
b) Latitude-velocity map of $^{13}$CO. 
The $^{13}$CO emission is integrated over the longitude from 24$\degr$.58 to 24$\degr$.92. 
c) The $^{13}$CO emission integrated over two different velocity ranges (i.e. 103--110.5 and 111--115 km s$^{-1}$) is presented, 
and the velocity ranges are also given in the figure. 
The contour levels of the background $^{13}$CO emission map are 10.5, 12, 15, 20, 25, 33, 40, 45, and 50 K km s$^{-1}$, 
while the broken contours (in red) are 4, 5.5, 8, 12, 16, 20, 23, 30, and 38 K km s$^{-1}$. 
d) Longitude-velocity map of $^{13}$CO. 
The $^{13}$CO emission is integrated over the latitude from $-$0$\degr$.23 to 0$\degr$.20. 
In both the left panels (i.e. Figures~\ref{sg7}a and~\ref{sg7}c), the positions of the 6.7 GHz masers are shown by triangles (in cyan). 
In both the right panels (i.e. Figures~\ref{sg7}b and~\ref{sg7}d), two velocities (i.e. 109 and 113 km s$^{-1}$) 
are highlighted by two dashed lines.}
\label{sg7}
\end{figure*}

To study the velocity field in the direction of the system N36, the integrated intensity map and position-velocity maps of $^{13}$CO are presented in Figure~\ref{sg7}.
For the comparison purpose, the integrated $^{13}$CO emission at [102, 116] km s$^{-1}$ is shown in Figure~\ref{sg7}a. 
Figures~\ref{sg7}b and~\ref{sg7}d display the latitude-velocity and longitude-velocity maps of $^{13}$CO, respectively.
In both the position-velocity maps, we have also highlighted two velocities (i.e. 109 and 113 km s$^{-1}$). These two velocity components appear to be linked through a relatively weak $^{13}$CO emission. In the position-velocity maps, these two clouds are separated by 4--12 km s$^{-1}$ in velocity.  
In Figure~\ref{sg7}c, we have shown the spatial distribution of the molecular gas in the 
two clouds at 103-110.5 km s$^{-1}$ and 111-115 km s$^{-1}$. Interestingly, the two clouds exhibit mutually overlapping areas (see also Figure~\ref{sg6}a).

\begin{figure*}
\epsscale{1.1}
\plotone{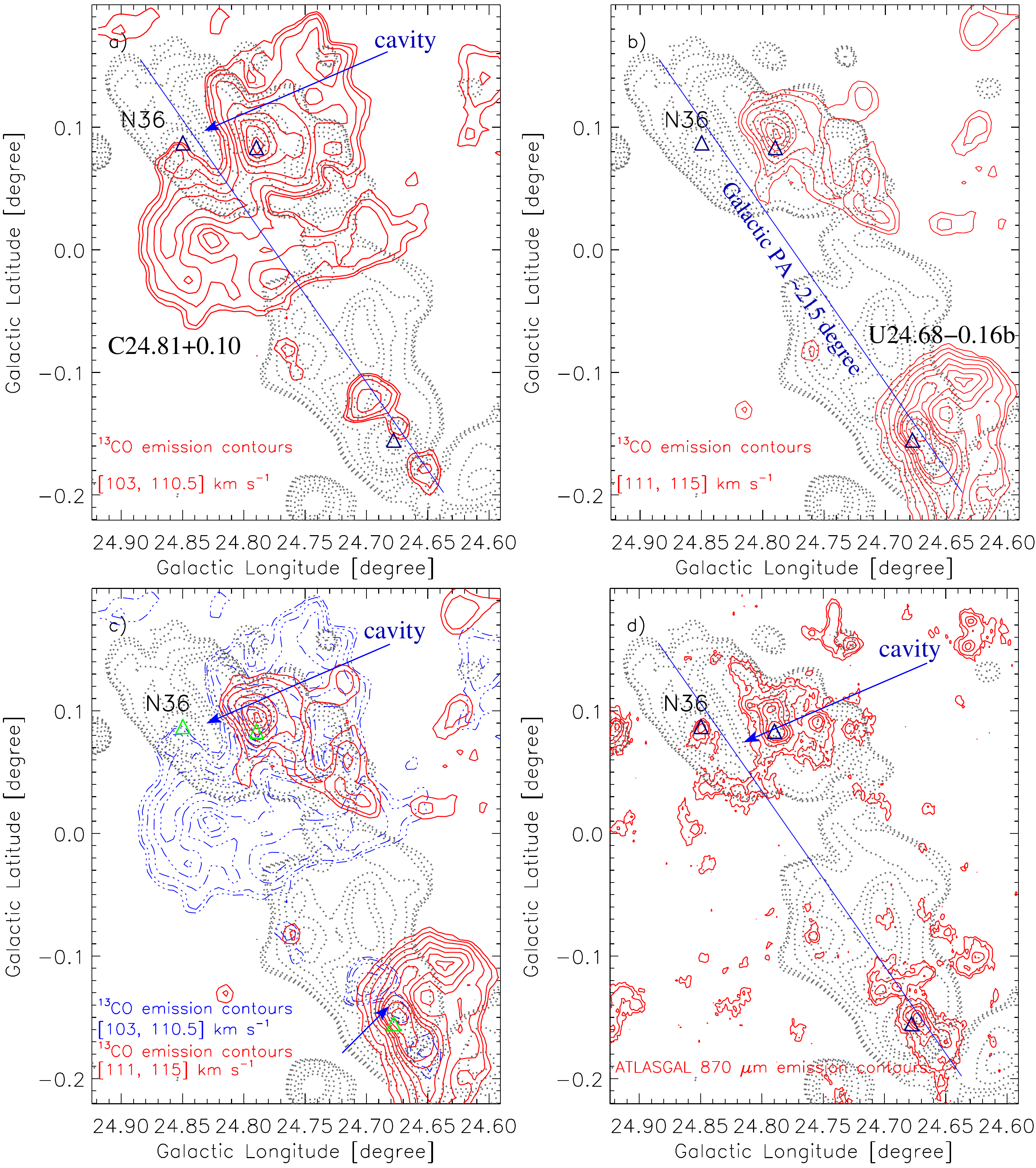}
\caption{a) Spatial distribution of the $^{13}$CO emission at [103, 110.5] km s$^{-1}$ and the NVSS 1.4 GHz continuum emission (see Figures~\ref{zfg1}a and~\ref{sg7}c). 
b) Spatial distribution of the $^{13}$CO emission at [111, 115] km s$^{-1}$ and the NVSS 1.4 GHz continuum emission (see Figures~\ref{zfg1}a and~\ref{sg7}c).
c) Spatial distribution of the $^{13}$CO emission integrated over two different velocity ranges (i.e. 103--110.5 and 111--115 km s$^{-1}$) and the NVSS 1.4 GHz continuum emission (see Figures~\ref{zfg1}a and~\ref{sg7}c). d) Spatial distribution of the continuum emission at ATLASGAL 870 $\mu$m and NVSS 1.4 GHz (see Figures~\ref{zfg1}a and~\ref{sg2}a). 
In three panels, a solid line is similar to as in Figure~\ref{zfg1}a. 
In all the panels, the positions of the 6.7 GHz masers are also shown by triangles.}
\label{sg8}
\end{figure*}

Figures~\ref{sg8}a and~\ref{sg8}b also show the spatial distribution of the molecular gas in the clouds at 103-110.5 and 111-115 km s$^{-1}$ against the NVSS 1.4 GHz continuum emission, respectively. 
One can notice that the majority of the molecular gas in the cloud C24.81+0.10 (at 103-110.5 km s$^{-1}$) is seen 
toward the MIR bubble N36, and a noticeable molecular emission is also distributed toward the H\,{\sc ii} 
region G24.68-0.16 (or cloud U24.68$-$0.16b). Similarly, a majority of the molecular gas in the cloud 
U24.68$-$0.16b (at 111-115 km s$^{-1}$) is found in the direction of the H\,{\sc ii} region G24.68-0.16, 
and a noticeable molecular gas is also traced toward the bubble N36 (or cloud C24.81+0.10). 
These results indicate that the clouds C24.81+0.10 and U24.68$-$0.16b are extended GMCs, and their central parts are severely affected by the ionized emission. 
This is the reason that there are no molecular and dust emissions found in the central parts of the clouds. 
In Figure~\ref{sg8}c, we have also displayed the spatial distribution of the $^{13}$CO emission integrated over two different velocity ranges (i.e. 103--110.5 and 111--115 km s$^{-1}$) against the NVSS 1.4 GHz radio continuum emission. 
Four H\,{\sc ii} regions (i.e. G24.80+0.10, G24.74+0.08, G24.71-0.13, and G24.68-0.16) are found in the common areas of the clouds. 
As mentioned earlier, the H\,{\sc ii} regions G24.80+0.10 and G24.68-0.16 contain the 6.7 GHz masers. 
Hence, the ongoing massive star formation activity is evident toward the common zones of the clouds. 
Figure~\ref{sg8}d shows the spatial distribution of the ATLASGAL 870 $\mu$m continuum emission against the NVSS 1.4 GHz radio continuum emission. 
A cavity-like feature is highlighted in the bubble N36, where the molecular and dust continuum emissions are absent (see Figures~\ref{sg8}a and~\ref{sg8}d). 
A solid line having the GPA of 215$\degr$ is also highlighted in 
Figures~\ref{sg8}a,~\ref{sg8}b, and~\ref{sg8}d. The cavity-like feature seen in the cloud at 103-110.5 km s$^{-1}$ appears remarkably parallel 
to the molecular cloud at 111-115 km s$^{-1}$ located in the southern direction (see arrows in Figure~\ref{sg8}c). 
These features are organized along the line with the GPA of 215$\degr$. Discussion on these results are presented in Section~\ref{sec:disc}.
\subsection{Embedded protostars and their distribution}
\label{subsec:phot1}
\begin{figure*}
\epsscale{0.71}
\plotone{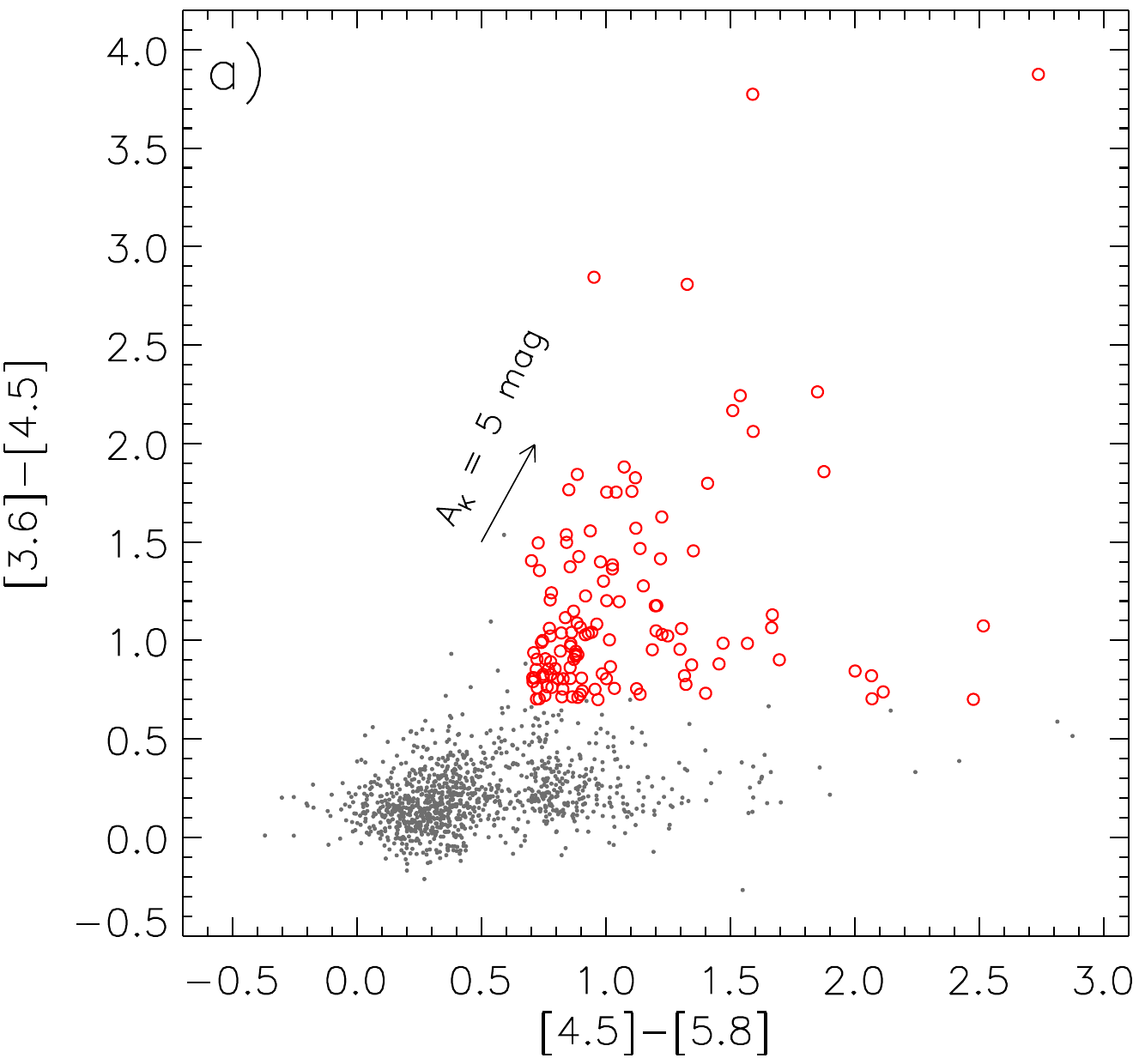}
\epsscale{1.1}
\plotone{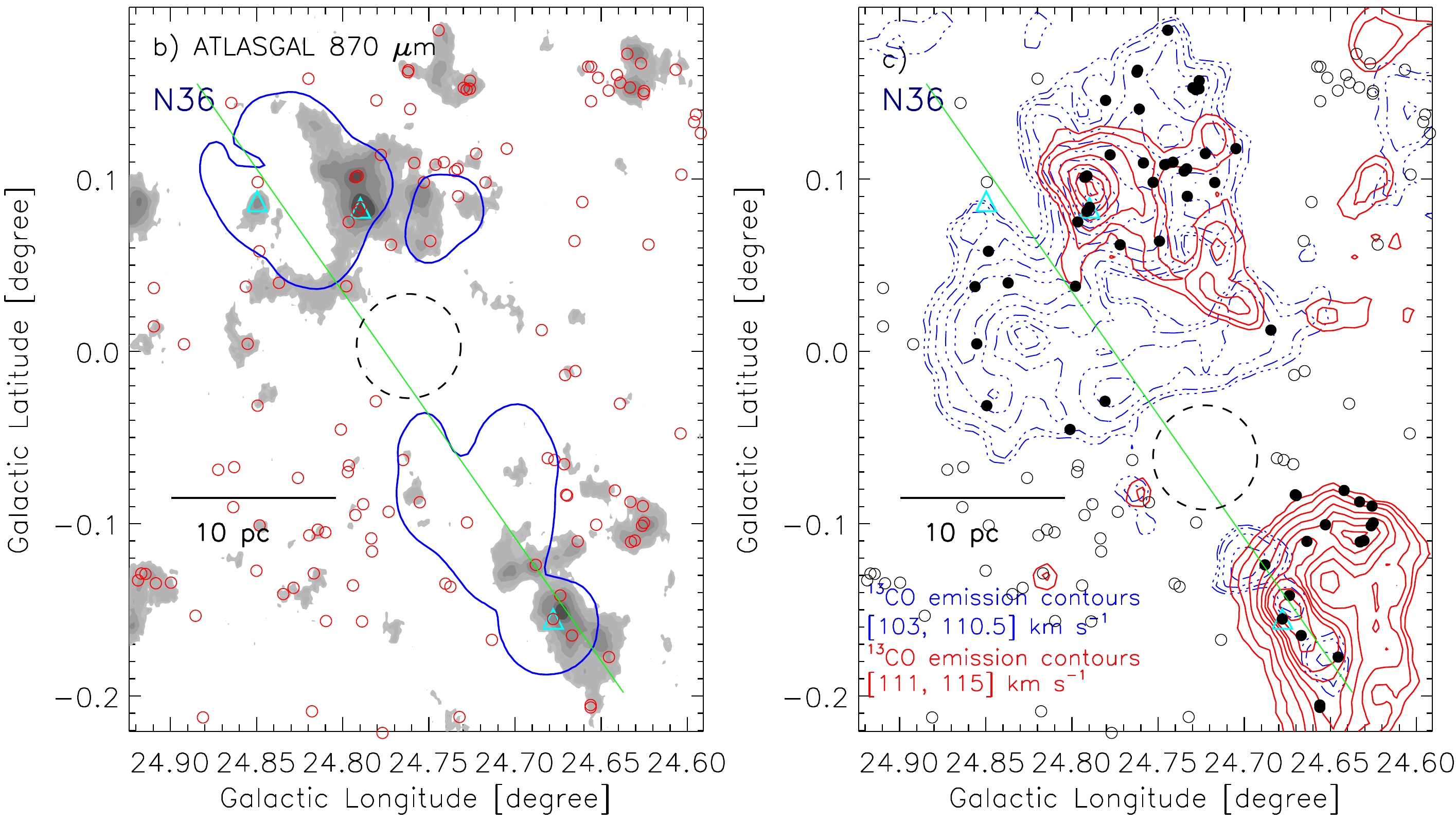}
\caption{a) Color-color plot ([3.6]$-$[4.5] vs. [4.5]$-$[5.8]) of sources observed in the our selected field. 
An extinction vector \citep[from][]{flaherty07} is shown in the figure. The dots (in gray) show the stars with only photospheric emission. 
Due to large numbers of stars with photospheric emission, we have plotted only some of these stars. 
The selected protostars are marked by red circles. 
b) Overlay of protostars (in red circles) on the ATLASGAL continuum map at 870 $\mu$m (see also Figure~\ref{sg2}a). 
The curves (in blue) are the same as in Figure~\ref{sg4}a. 
c) Overlay of protostars (see circles) on the molecular intensity maps. The maps are the same as in Figure~\ref{sg8}c. 
The protostars seen inside the clouds are shown by the filled circles (in black), while the protostars lying outside the clouds are represented by open circles (in black). In the panels ``b" and ``c", the positions of the 6.7 GHz masers are also shown by triangles, 
and a dashed big circle indicates the gas-devoid area. 
In the panels ``b" and ``c", a solid line is similar to as in Figure~\ref{zfg1}a.}
\label{sg9}
\end{figure*}

This section deals with the selection of the embedded protostars in our selected target field. 
To detect infrared excess emission from the protostars, 
we have employed the {\it Spitzer} photometric data. 
To select protostars, \citet{hartmann05} and \citet{getman07} explored the color-color ([4.5]$-$[5.8] vs [3.6]$-$[4.5]) space using the {\it Spitzer} 3.6, 4.5, and 5.8 photometric data, and provided the selection conditions (i.e. [4.5]$-$[5.8] $\ge$ 0.7 mag and [3.6]$-$[4.5] $\ge$ 0.7 mag). Figure~\ref{sg9}a shows a color-color plot ([4.5]$-$[5.8] vs [3.6]$-$[4.5]) of point-like sources. Based on the color conditions, we have identified 125 protostars, which are shown by red circles in Figure~\ref{sg9}a. 
There is also possibility that our selected protostars might be contaminated by asymptotic giant branch (AGB) stars. 
Using the {\it Spitzer} 4.5--24 $\mu$m bands, \citet{robitaille08} proposed a condition (i.e. [4.5] $>$ 7.8 mag and [8.0]-[24.0] $<$ 2.5 mag) to infer possible AGB contaminants. 
Based on the availability of the MIPSGAL photometric data at 24 $\mu$m \citep[e.g.,][]{gutermuth15} for our selected protostars, we followed the work of \citet{robitaille08} and tried to identify the possible AGB contaminants. We find that our selected protostars appear to be free from the AGB contaminants. 

Figures~\ref{sg9}b and~\ref{sg9}c display the overlay of the selected protostars (see circles) on the ATLASGAL 870 $\mu$m continuum contour map and the molecular intensity maps, respectively. 
Figure~\ref{sg9}b reveals the spatial correlation between protostars and cold dust emission, tracing the ongoing star formation activity toward the dense regions in the system N36. The protostars, dust emission, and molecular emission are not detected between the northern and southern H\,{\sc ii} regions (see a dashed big circle in Figures~\ref{sg9}b and~\ref{sg9}c). 
In Figure~\ref{sg9}c, the filled circles and open circles indicate the distribution of protostars inside and outside the molecular clouds, respectively.
\section{Discussion}
\label{sec:disc}
\subsection{Impact of massive stars in the system N36}
\label{subsecx:dd1}
In the present work, we have selected a configuration of an extended physical system of O-type stars (i.e. system N36) 
toward {\it l} = 24$\degr$.8, b = 0$\degr$.1. The system is prominently seen in the {\it Herschel} temperature and NVSS 1.4 GHz continuum maps (see Figures~\ref{zfg1}a and~\ref{sg3}a). 
The existence of the extended temperature structure in the system gives a clue of the feedback from massive stars (such as, stellar wind, 
ultraviolet radiation, and pressure-driven H\,{\sc ii} region) powering the H\,{\sc ii} regions.
The molecular and dust materials are not observed toward the central parts of the extended clouds C24.81+0.10 and U24.68$-$0.16b, 
which probably indicates that the H\,{\sc ii} regions in the system are interacting with their immediate environment. 
In the system, several massive clumps ($>$ 10$^{3}$ M$_{\odot}$) are found. 
Several embedded protostars \citep[average age $\sim$0.44 Myr;][]{evans09} distributed toward the 
clumps are also identified in the system N36, depicting the ongoing star formation activity. 
Three 6.7 GHz methanol masers are observed in the system, suggesting the presence of 
early phases of massive star formation ($<$ 0.1 Myr). 
The dynamical ages of the H\,{\sc ii} regions in the system vary between 0.4 and 1.3 Myr (see Section~\ref{subsec:radio}).
Hence, it is possible that some of the H\,{\sc ii} regions in the system N36 are old enough 
to influence the formation of embedded protostars \citep[e.g.,][]{elmegreen98}, 
which cannot be neglected in the system  N36 \citep[e.g.,][]{deharveng10}.

However, it remains unknown how the system of O-type stars is formed, and which process can be responsible for 
the observed velocity separation between two clouds (i.e. 4--12 km s$^{-1}$) seen in the direction of the system N36. 
To assess the role of the feedback of massive star(s) for explaining the observed velocity separation, we have computed the mechanical 
energy (E$_{w}$ $\sim$3 $\times$ 10$^{48}$ ergs) that can be liberated out by the massive O6V star in a given timescale (i.e. 0.5 Myr). 
In the analysis, we have first estimated the expected mechanical luminosity of the stellar wind (L$_{w}$ = 0.5$\, \dot{M}_{w}$ V$_{w}^{2}$ erg s$^{-1}$) for the O6V star where 
the mass-loss rate \citep[$\dot{M}_{w}$ = 2.0 $\times$ 10$^{-7}$ M$_{\odot}$ yr$^{-1}$;][]{dejager88} and the wind velocity \citep[V$_{w}$ = 2500 km s$^{-1}$;][]{prinja90} are considered. In the next step, we have compared the value of E$_{w}$ with the kinematic energy ($\sim$7.3 $\times$ 10$^{48}$ ergs) of the cloud C24.81+0.10 (M$_{cloud}$ $\sim$45385 M$_{\odot}$) at a velocity of 4 km s$^{-1}$. 
We have also computed the kinematic energy (i.e. $\sim$2.7 $\times$ 10$^{48}$ ergs) of the cloud at 111-115 km s$^{-1}$ (M$_{cloud}$ $\sim$16975 M$_{\odot}$; see Figure~\ref{sg8}b) with a velocity of 4 km s$^{-1}$, which is also comparable to the value of E$_{w}$. 
These results are also valid for a velocity separation of 12 km s$^{-1}$. 
Hence, we find that the stellar feedback is even more inefficient for a velocity separation larger than 4 km s$^{-1}$. 
On the basis of these calculations, we find that the stellar feedback cannot explain the velocity separation between two clouds. 
It also implies that the feedback of massive stars does not explain the synchronous birth of O stars in the system N36. 
For estimating the molecular masses of the clouds, we have followed the work of \citet{yan16} (see also equations 4 and 5 in their paper). 
\subsection{Location of the system N36 in the Milky Way}
\label{subsecx:dd2}
The location of the system of H\,{\sc ii} regions hosting the bubble N36 in the Milky Way is inferred through the knowledge of its kinematic distance (i.e. 6 kpc), which has been derived with the help of the Galactic rotation curve \citep[e.g.,][]{mcclure07,reid09,reid14} and the velocity measurements of 
the H\,{\sc i}/CO/6.7 GHz maser/ionized gas \citep[e.g.,][]{anderson09,hou14,hu16,urquhart18}. 
However, one obtains two possible kinematic distances corresponding to one V$_{lsr}$ in the inner Galaxy, referred to as kinematic distance ambiguity. 
In earlier studies, the near-kinematic distance (i.e. $\sim$6 kpc) to the H\,{\sc ii} regions in the system of O-type stars was adopted \citep[see] [for more details]{kantharia07,anderson09,jones12}. \citet{hu16} reported two 6.7 GHz methanol masers toward the bubble N36 at a distance of $\sim$9 kpc, while \citet{hou14} listed a distance of $\sim$6 kpc
for one of the 6.7 GHz methanol masers reported by \citet{hu16}. Furthermore, \citet{hu16} also derived a distance of $\sim$6 kpc for another 6.7 GHz methanol maser in the system of H\,{\sc ii} regions. 
Based on the extensive work of \citet{urquhart18}, the kinematic distance ambiguity of the sources in our selected target field has also been resolved. 

\citet{hou14} studied the distribution of the H\,{\sc ii} regions against 
a spiral arm model (see Figure~5 in their paper). They showed the locations of H\,{\sc ii} regions, different spiral arms, and the Galactic bar in their Figure~5. 
Following their work, the adopted distance to the system N36 (i.e. d = 6.0 kpc) favours its location ({\it l} = 24$\degr$.6 -- 24$\degr$.9) 
in the direction of the interface of the Galactic bar and the Norma Galactic arm in the Milky Way \citep[see lower right panel in Figure~5 in][]{hou14}.
\subsection{Star formation scenario in the system N36}
\label{subsecx:dd3}
In Section~\ref{sec:intro}, based on the knowledge of V$_{lsr}$ of the ATLASGAL clumps, we have identified 
at least three different velocity components (i.e. [80, 100], [104, 113], and [115, 122] km s$^{-1}$) in the direction 
of {\it l} = 23$\degr$.9 -- 24$\degr$.9 (see Figure~\ref{ffg1}a). 
In the velocity space of $^{13}$CO, these velocity components are also seen in this selected longitude direction (see Figures~\ref{ffg1}b and~\ref{ffg1}c). 
The existence of different velocity components further supports that the system of O-type stars is located in the direction of the bar-arm interface \citep[e.g.,][]{beuther17}.  
In the bar-arm interface, there is possibility of the collisions of molecular clouds due to the bar potential. 
It has also been suggested that colliding streams of gas from the bar and the arm may favour increased star formation \citep[e.g.,][]{nguyen11,beuther17}. 

In recent years, the cloud-cloud collision (CCC) process is being investigated to explain the observed star formation activities at the junction of molecular clouds or the shock-compressed interface \citep[e.g.,][and references therein]{habe92,anathpindika10,inoue13,takahira14,haworth15a,haworth15b,torii17,bisbas17,balfour17,takahira18}. 
In the CCC process, the maximum star-formation is expected close to the point of impact or collision interface \citep{haworth15a}. 
The spatial and velocity connection of two clouds with a large velocity separation provides a promising hint of the CCC process \citep[e.g.,][]{furukawa09,ohama10,ohama17b,ohama17,fukui14,fukui16,fukui18a,fukui18,baug16,dewangan17x,dewangan17,
dewangan17b,dewangan17xx,dewangan18b,fujita17,torii17,torii17x,sano18}. 
The detection of complementary distributions of clouds is considered as an another important evidence of the CCC process \citep[e.g.][]{fukui18a}. In this context, a cavity exists in one of the molecular clouds, and is referred to as ``Keyhole" feature or an intensity depression in the 
molecular distribution. Additionally, the other cloud is referred to as ``Key" feature or intensity enhancement, which can be displaced with respect to the ``Keyhole" feature \citep[see][for more details]{fukui18a}. These authors utilized a method to estimate the displacement by optimizing the overlap of the intensity enhancement and depression.   

\begin{figure*}
\epsscale{1}
\plotone{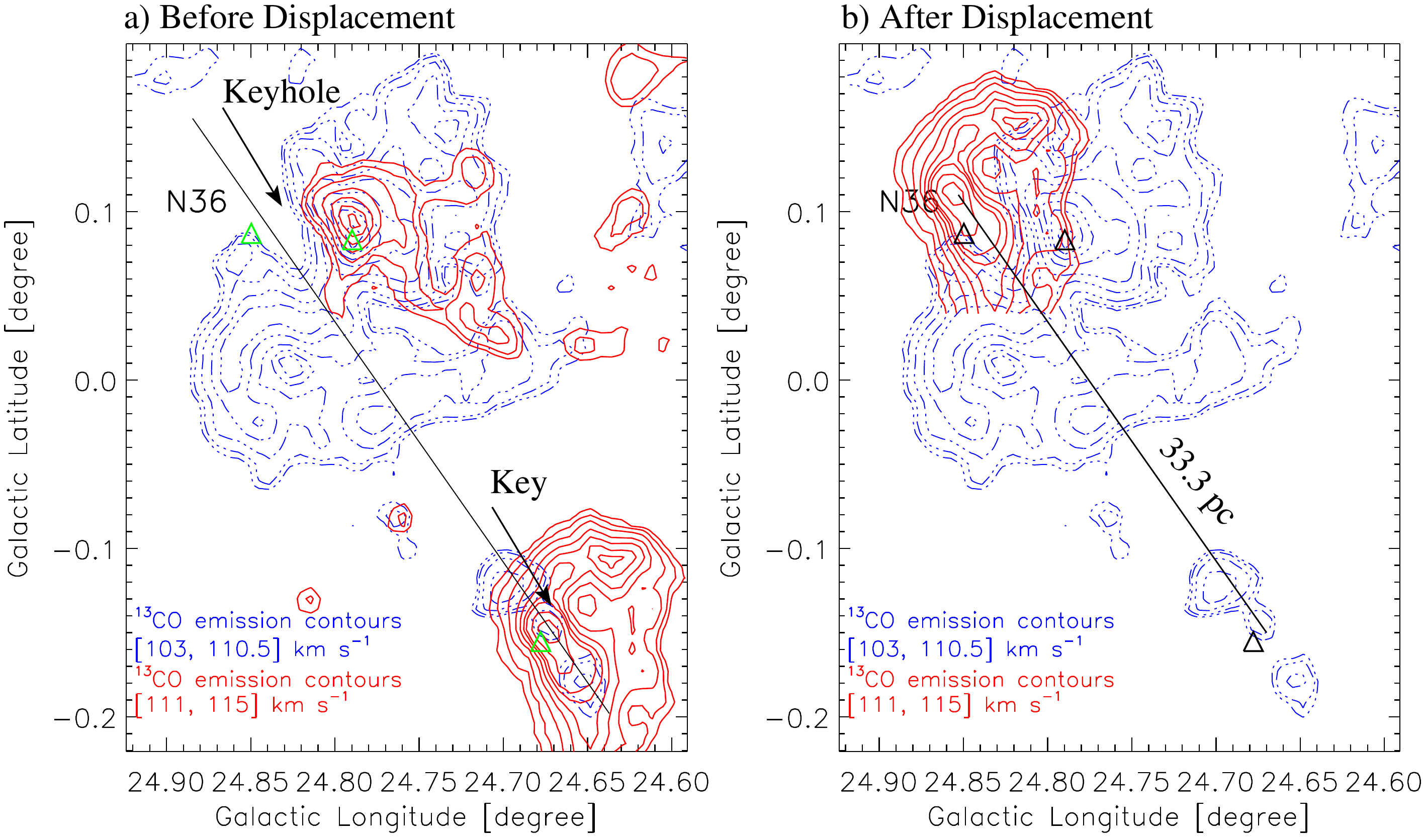}
\caption{a) Spatial distribution of the molecular gas associated with clouds at [103, 110.5] and [111, 115] km s$^{-1}$, similar to as in Figure~\ref{sg8}c. 
b) Same as Figure~\ref{zzsg9}a, but the cloud component at 111--115 km s$^{-1}$ is displaced in the northern direction along a line having the GPA of 215$\degr$.}
\label{zzsg9}
\end{figure*}

In Section~\ref{sec:coem}, we have found that there are two molecular cloud components (at 103-110.5 and 111-115 km s$^{-1}$) 
in the system, which are linked in the velocity space (see the velocity channel maps and position-velocity plots). 
A spatial separation between the northern edge of U24.68$-$0.16b (at 111-115 km s$^{-1}$) 
and the southern edge of C24.81+0.10 (at 103-110.5 km s$^{-1}$) can also be seen in Figure~\ref{zzsg9}a. 
In Section~\ref{subsecx:dd1}, our calculations suggest that the cloud C24.81+0.10 (at 103-110.5 km s$^{-1}$) is more massive compared to the cloud U24.68$-$0.16b (at 111-115 km s$^{-1}$). 
A careful examination of the molecular gas at 103-110.5 km s$^{-1}$ in the system N36, we find a cavity or ``Keyhole" configuration in the northern direction along a line with the GPA of 215$\degr$ (see Figure~\ref{zzsg9}a).
Additionally, a displaced ``Key" feature, traced at 111-115 km s$^{-1}$, is identified in the southern direction, and appears remarkably parallel to the ``Keyhole" configuration (see Figure~\ref{zzsg9}a). 
In Figure~\ref{zzsg9}b, we have applied a displacement of $\sim$33.3 pc to the cloud at 111-115 km s$^{-1}$ 
in the northern direction along a line having the GPA of 215$\degr$ (see a solid line in Figure~\ref{zzsg9}b).
This exercise gives a complementary fit between the clouds at 103-110.5 and 111-115 km s$^{-1}$, 
which indicates the interaction of these two clouds in the system N36. 
The distribution of the dust emission, molecular gas, and protostars is not found between these clouds (or ``Key" and ``Keyhole" features), where only ionized emission is observed. The ionized emission is also aligned with the ``Key" and ``Keyhole" features. Altogether, our results support the existence of interacting molecular clouds in the system N36. 
We have also traced reliable signatures of massive star formation toward the common zones of the clouds (i.e., C24.81+0.10 and U24.68$-$0.16b). 
Hence, the CCC scenario (i.e., the collision between the cloud components around 109 and 113 km s$^{-1}$) is applicable in the system, which may explain the formation of the H\,{\sc ii} regions in the system. 
The lifetime of the collision of molecular clouds is computed to be $\sim$2.7--8.2 Myr. 
The ages of the H\,{\sc ii} regions are estimated to be about 0.4--1.3 Myr. These calculations also indicate the onset of the CCC process. 
In the calculation, we have adopted a viewing angle (= 45$\degr$) of the collision to the line of sight.
Hence, the distance between the two clouds is considered to be $\sim$47 pc (= 33.3 pc/sin(45$\degr$)). 
The velocity difference estimated from the observed relative velocity is found to be $\sim$5.6--17 km s$^{-1}$ (= 4--12/cos(45$\degr$) 
(in km s$^{-1}$); see Section~\ref{sec:coem}). 
The lower end in the estimated lifetime of the collision interval (i.e., $\sim$2.7 Myr) corresponding to the higher end of the velocity range (i.e., 12 $\times$ $\sqrt{2}$ km s$^{-1}$) is more consistent with the ages of the H\,{\sc ii} regions, suggesting a recent onset of massive star formation.  
Considering the location of the system N36 in the inner Galaxy, optically thin molecular line 
observations will provide an opportunity to explore further the collision scenario in the system N36. 

As mentioned in Section~\ref{sec:intro}, in the direction of {\it l} = 23$\degr$.9 -- 24$\degr$.9, a cluster of the ATLASGAL clumps is found toward the MIR bubble N35, and these clumps are situated at a distance of 7.8 kpc. \citet{torii17x} carried out an observational study of the GMC associated with the bubble N35 and two H\,{\sc ii} regions (i.e., G024.392+00.072, and G024.510-00.060), and adopted a distance of the GMC to 8.8 kpc. They proposed a CCC scenario (i.e., the collisions between the lower-velocity component (LVC) at 110--114 km s$^{-1}$ and three higher-velocity-components (HVCs) at 118--126 km s$^{-1}$) to explain the formation of the bubble N35 and two H\,{\sc ii} regions. 
They discussed that the collisions between molecular clouds had begun since less than $\sim$1 Myr 
ago. Considering the distances of the bubble N35 and the system N36, our selected target system does not have any physical connection with the bubble N35. However, the CCC scenario in both the systems has been proposed, and 
the collisions took place in both the systems independently. 
\section{Summary and Conclusions}
\label{sec:conc}
In order to investigate a large-scale system of several MIR bubbles/H\,{\sc ii} regions 
hosting the O-type stars and the birth of such extended system, we present the multi-wavelength data 
analysis of a field (size $\sim$0$\degr$.33 $\times$ 0$\degr$.42) containing several H\,{\sc ii} regions 
toward {\it l} = 24$\degr$.6 -- 24$\degr$.9, {\it b} = $-$0$\degr$.2 -- 0$\degr$.2.\\\\ 
$\bullet$ Using the existing catalog of the ATLASGAL dust continuum clumps at 870 $\mu$m and the NVSS 1.4 GHz continuum map, 
a configuration of at least four H\,{\sc ii} regions (i.e. ``system N36"; extension $\sim$35 pc) is observationally found in the inner Galaxy 
at a distance of 6.0 kpc. These H\,{\sc ii} regions are G24.83+0.10 (containing two compact/ultra-compact H\,{\sc ii} 
regions G24.85+0.09 and G24.80+0.10), G24.74+0.08, G24.71-0.13, and G24.68-0.16.\\ 
$\bullet$ Each of the H\,{\sc ii} regions in the system is powered by a radio spectral type of O star. 
The dynamical ages of the H\,{\sc ii} regions vary between 0.4 and 1.3 Myr for 10$^{3}$ cm$^{-3}$ ambient density.\\
$\bullet$ At least one 6.7 GHz methanol maser is detected toward three H\,{\sc ii} regions (i.e. G24.85+0.09, 
G24.80+0.10, and G24.68-0.16) in the system. \\    
$\bullet$ The system N36 ({\it l} = 24$\degr$.6 -- 24$\degr$.9; d = 6.0 kpc) is located in the Milky Way where the 
Galactic bar meets the Norma Galactic arm (i.e., the bar-arm interface).\\ 
$\bullet$ In the {\it Herschel} temperature map, the extended temperature structure is observed toward the system of H\,{\sc ii} regions 
which is well traced in a temperature range of 23 to 30 K.\\
$\bullet$ Twenty seven ATLASGAL clumps are found 
in the system. Several clumps are identified with high bolometric 
luminosity ($>$ 10$^{3}$ L$_{\odot}$), and are massive ($>$ 10$^{3}$ M$_{\odot}$).\\
$\bullet$ Using the {\it Spitzer} color-color ([4.5]$-$[5.8] vs [3.6]$-$[4.5]) space, 125 embedded protostars are identified in our selected target field, and majority of them are found toward the ATLASGAL dust clumps in the system. \\
$\bullet$ Using the GRS $^{13}$CO line data, the molecular gas associated with the system is 
studied in a velocity range of 102--116 km s$^{-1}$.\\ 
$\bullet$ The spatial distribution of the dust emission, molecular gas, and ionized emission suggests an impact of massive stars powering the 
H\,{\sc ii} regions in the system N36.\\ 
$\bullet$ The $^{13}$CO line data reveal two extended 
GMCs (i.e. C24.81+0.10 (around 109 km s$^{-1}$) and U24.68$-$0.16b (around 113 km s$^{-1}$)) 
in the direction of the system N36. These two clouds are interconnected in the velocity space, and are separated by 4--12 km s$^{-1}$ in velocity.\\
$\bullet$ A ``Keyhole" feature in the cloud at 103-110.5 km s$^{-1}$ (i.e. C24.81+0.10) and a ``Key" feature in the cloud at 111-115 
km s$^{-1}$ (i.e. U24.68$-$0.16b) are identified in the system N36. A displacement of $\sim$33.3 pc enables a spatial fit between the Keyhole and Key features.\\
$\bullet$ Signposts of massive star formation (i.e. H\,{\sc ii} regions and 6.7 GHz methanol masers) are investigated at the common zones of the two clouds, where the embedded protostars are also traced.\\ 
$\bullet$ The observational results suggest the onset of the CCC process in the system N36. 
Adopting the collision at a viewing angle 45$\degr$ to the line of sight, the lifetime of the collision is estimated to be $\sim$2.7--8.2 Myr.\\

Taking into account all the observational results, 
the existence of the H\,{\sc ii} regions in the system seems to be explained by the interaction of molecular 
clouds at the bar-arm interface. 
In the direction of the system N36, high-resolution and optically thin molecular line data 
will be useful to further gain more insights in the CCC process. 
\acknowledgments 
We thank the anonymous reviewer for a critical reading of the manuscript and several useful comments and 
suggestions, which greatly improved the scientific contents of the paper. 
The research work at Physical Research Laboratory is funded by the Department of Space, Government of India. 
This publication made use of data products from the the Spitzer Space Telescope, which is operated by the Jet Propulsion Laboratory, California Institute of Technology under a contract with NASA.
This publication makes use of molecular line data from the Boston University-FCRAO Galactic
Ring Survey (GRS). The GRS is a joint project of Boston University and Five College Radio Astronomy Observatory, 
funded by the National Science Foundation (NSF) under grants AST-9800334, AST-0098562, and AST-0100793. 
The National Radio Astronomy Observatory is a facility of the National Science Foundation operated under cooperative agreement by Associated Universities, Inc. 
IZ is supported by the Russian Foundation for Basic Research (RFBR) grants No. 17-52-45020 and 18-02-00660. 
%
%

%
%
%
\begin{table*}
\setlength{\tabcolsep}{0.1in}
\centering
\caption{Physical parameters of the ATLASGAL dust clumps at 870 $\mu$m (see Figure~\ref{sg1}a). These parameters are taken from \citet{urquhart18}. 
Table lists ID, Galactic coordinates ({\it l}, {\it b}), 870 $\mu$m peak flux density (P$_{870}$), 
870 $\mu$m integrated flux density (S$_{870}$), radial velocity (V$_{lsr}$), distance, clump effective radius ($R_\mathrm{c}$), dust temperature ($T_\mathrm{d}$), 
bolometric luminosity ($L_{bol}$), clump mass ($M_\mathrm{clump}$), and H$_{2}$ column density ($N(\mathrm H_2)$).} 
\label{tab2}
\begin{tabular}{lcccccccccccr}
\hline 
  ID  &  {\it l}     &  {\it b}    & P$_{870}$   & S$_{870}$  & V$_{lsr}$ &distance  &R$_\mathrm{c}$ &$T_\mathrm{d}$ &$logL_{bol}$ &log$M_\mathrm{clump}$&log$N(\mathrm H_2)$\\  
      & (degree) & (degree) &  (Jy/beam)  &  (Jy)      & (km s$^{-1}$)   & (kpc)   &     (pc)     & (K)&   ($L_\odot$)& ($M_\odot$) &(cm$^{-2}$)\\  
\hline
\hline 
c1  & 24.624 & -0.101 &  1.34 &    9.07 &  113.3 &    6.0  &   2.93 &	18.3 &    3.653 &    3.319 &   22.584 \\
c2  & 24.633 & -0.109 &  0.86 &    5.06 &  110.7 &    6.0  &   1.61 &	14.0 &    2.988 &    3.252 &   22.578 \\
c3  & 24.651 & -0.169 &  1.33 &    13.28 &  113.0 &    6.0  &	3.21 &   19.0 &    4.047 &    3.460 &	22.556 \\
c4  & 24.654 & -0.204 &  0.45 &    3.24 &  110.2 &    6.0  &   0.70 &	20.6 &    3.561 &    2.796 &   22.033 \\
c5  & 24.659 & -0.186 &  0.58 &    2.62 &  109.8 &    6.0  &   0.70 &	22.4 &    3.523 &    2.651 &   22.091 \\
c6  & 24.666 & -0.116 &  0.86 &    7.22 &  112.2 &    6.0  &   1.68 &	18.8 &    3.506 &    3.202 &   22.373 \\
c7  & 24.673 & -0.151 &  4.83 &    39.11 &  112.7 &    6.0  &	4.82 &   25.7 &    5.098 &    3.742 &	22.929 \\
c8  & 24.689 & -0.126 &  1.32 &    7.63 &  108.2 &    6.0  &   1.61 &	14.2 &    2.841 &    3.420 &   22.753 \\
c9  & 24.703 & -0.127 &  1.14 &    6.26 &  108.6 &    6.0  &   1.88 &	21.6 &    3.823 &    3.052 &   22.407 \\
c10 & 24.709 & -0.117 &  0.51 &    3.09 &  104.8 &    6.0  &   0.70 &	27.2 &    3.713 &    2.607 &   21.920 \\
c11 & 24.728 &  0.086 &  0.49 &    3.89 &  107.8 &    6.0  &   0.84 &	25.2 &    3.443 &    2.750 &   21.947 \\
c12 & 24.728 &  0.152 &  1.32 &    4.63 &  108.9 &    6.0  &   2.16 &	25.0 &    4.137 &    2.831 &   22.382 \\
c13 & 24.741 &  0.182 &  0.66 &    2.71 &  108.9 &    6.0  &   1.40 &	20.9 &    3.489 &    2.707 &   22.190 \\
c14 & 24.744 &  0.162 &  0.78 &    5.44 &  107.4 &    6.0  &   2.30 &	21.7 &    3.795 &    2.987 &   22.240 \\
c15 & 24.754 &  0.061 &  0.71 &    6.09 &  109.9 &    6.0  &   1.61 &	26.5 &    3.893 &    2.915 &   22.078 \\
c16 & 24.754 &  0.091 &  2.09 &    15.45 &  109.4 &    6.0  &	2.37 &   23.8 &    4.195 &    3.383 &	22.611 \\
c17 & 24.759 &  0.164 &  0.49 &    1.56 &  107.5 &    6.0  &   0.70 &	16.9 &    2.648 &    2.606 &   22.200 \\
c18 & 24.759 & -0.084 &  0.57 &    1.58 &  110.8 &    6.0  &   0.70 &	24.5 &    4.231 &    2.377 &   22.029 \\
c19 & 24.789 &  0.082 & 15.52 &    56.56 &  110.2 &    6.0  &	2.93 &   26.5 &    5.173 &    3.883 &	23.418 \\
c20 & 24.796 &  0.101 &  3.52 &    39.86 &  109.9 &    6.0  &	4.40 &   28.1 &    5.423 &    3.697 &	22.740 \\
c21 & 24.806 &  0.039 &  0.64 &    10.23 &  106.2 &    6.0  &	2.58 &   23.5 &    3.567 &    3.212 &	22.105 \\
c22 & 24.809 &  0.104 &  1.55 &    1.63 &  109.7 &    6.0  &   0.70 &	30.2 &    4.285 &    2.268 &   22.343 \\
c23 & 24.816 &  0.131 &  0.75 &    7.11 &  109.5 &    6.0  &   3.35 &	34.3 &    5.053 &    2.837 &   21.957 \\
c24 & 24.818 &  0.036 &  0.54 &    7.24 &  108.1 &    6.0  &   1.54 &	26.9 &    3.567 &    2.982 &   21.951 \\
c25 & 24.844 &  0.036 &  0.58 &    6.06 &  106.7 &    6.0  &   1.68 &	21.7 &    3.553 &    3.033 &   22.111 \\
c26 & 24.849 &  0.086 &  1.19 &    4.16 &  109.0 &    6.0  &   1.40 &	33.9 &    4.622 &    2.611 &   22.164 \\
c27 & 24.854 &  0.004 &  0.57 &    2.95 &  105.0 &    6.0  &   1.68 &	16.3 &    2.847 &    2.908 &   22.290 \\
\hline          
\end{tabular}
\end{table*}
\begin{table*}
\setlength{\tabcolsep}{0.1in}
\centering
\caption{Physical parameters of four H\,{\sc ii} regions (see Figure~\ref{sg3}b). Table lists ID, Galactic coordinates ({\it l}, {\it b}), 
deconvolved effective radius of the H\,{\sc ii} region ($R_\mathrm{HII}$), total flux (S${_\nu}$), 
Lyman continuum photons (log$N_\mathrm{uv}$), dynamical age ($t_\mathrm{dyn}$), radio spectral type, electron density (n$_{e}$), emission measure (EM), and mass of ionized hydrogen (M$_{HII}$). 
Note that the extended H\,{\sc ii} region G24.83+0.10 hosts two compact/ultra-compact H\,{\sc ii} regions G24.85+0.09 and G24.80+0.10.} 
\label{tab3}
\begin{tabular}{lcccccccccccr}
\hline 
  ID  &  {\it l}     &  {\it b}    &  $R_\mathrm{HII}$   & S${_\nu}$  & log$N_\mathrm{uv}$  &t$_{dyn}$  &Spectral Type &n$_{e}$ &EM &M$_{HII}$&Other\\  
      & (degree) & (degree) &  (pc)               &  (Jy)      &  (s$^{-1}$)           & (Myr)   &     (V)     & (cm$^{-3}$)&(cm$^{-6}$ pc)&($M_\odot$)&names\\  
\hline
\hline 
   1&	  24.800  &    0.095	&  5.05 &    5.07  &  49.16  &   1.02&	   O6.5-O6&59 &33160 &785 & G24.83+0.10  \\
   2&	  24.745 &     0.075&	  2.28  &   0.42  &  48.08  &   0.47&	  O9.5    &56 &13450 &70 & G24.74+0.08  \\
   3&	  24.716 &    -0.130&	  5.07  &   1.95 &   48.74 &    1.32&	  O7.5-O7 &36 &12620 &490 & G24.71-0.13  \\
   4&	  24.679&     -0.163&	  2.66  &   1.89 &   48.73 &    0.41&	 O7.5-O7  &94 &44450 &184 & G24.68-0.16  \\
\hline          
\end{tabular}
\end{table*}


\begin{thebibliography}{}
%
\bibitem[Anathpindika(2010)]{anathpindika10}
Anathpindika, S.~V. 2010, MNRAS, 405, 1431

\bibitem[Anderson \& Bania (2009)]{anderson09}
Anderson, L.~D. \& Bania, T.~M. 2009, ApJ, 690, 706

\bibitem[Anderson et al.(2009)]{anderson09a}
Anderson, L.~D., Bania, T.~M., Jackson, J.~M., et al. 2009, ApJS, 181, 255

\bibitem[Balfour et al.(2017)]{balfour17}
Balfour, S.~K., Whitworth, A.~P., \& Hubber, D.~A. 2017, MNRAS, 465, 3483

\bibitem[Baug et al.(2016)]{baug16}
Baug, T., Dewangan, L.~K., Ojha, D.~K., \& Ninan, J.~P. 2016, ApJ, 833, 85

\bibitem[Benjamin et al.(2003)]{benjamin03}
Benjamin, R.~A.,Churchwell, E., Babler, B.~L., et al. 2003, PASP, 115, 953

\bibitem[Benjamin et al.(2005)]{benjamin05}
Benjamin, R.~A.,Churchwell, E., Babler, B.~L., et al. 2005, ApJ, 630, 149

\bibitem[Beuther et al.(2017)]{beuther17}
Beuther, H., Meidt, S., Schinnerer, E., Paladino, R., \& Leroy, A. 2017, A\&A, 597, 85

\bibitem[Bisbas et al.(2017)]{bisbas17}
Bisbas, T.~G., Tanaka, K.~E.~I., Tan, J.~C., Wu, B., \& Nakamura, F. 2017, ApJ, 850, 23

\bibitem[Bronfman(2008)]{bronfman08}
Bronfman, L. 2008, Ap. Space Sci., 313, 81-85

\bibitem[Carey et al.(2005)]{carey05}
Carey, S. J., Noriega-Crespo, A., Price, S.~D., et al. 2005, BAAS, 37, 1252

\bibitem[Churchwell et al.(2006)]{churchwell06}
Churchwell, E., Povich, M.~S., Allen, D., et al. 2006, ApJ, 649, 759

\bibitem[Churchwell et al.(2007)]{churchwell07}
Churchwell, E., Watson, D.~F., Povich, M.~S., et al. 2007, ApJ, 670, 428

\bibitem[Condon et al.(1998)]{condon98}
Condon, J.~J., Cotton, W.~D., Greisen, E.~W., et al. 1998, AJ, 115, 1693

\bibitem[Deharveng et al.(2010)]{deharveng10}
Deharveng, L., Schuller, F., Anderson, L.~D., et al. 2010, A\&A, 523, 6

\bibitem[de Jager et al.(1988)]{dejager88}	
de Jager, C., Nieuwenhuijzen, H., \&  van der Hucht, K. A. 1988, A\&AS, 72,259

\bibitem[Dewangan et al.(2015a)]{dewangan15a}
Dewangan, L.~K., Ojha, D.~K., Grave, J.~M.~C., \& Mallick, K.~K. 2015a, MNRAS, 446, 2640
%
\bibitem[Dewangan et al.(2015b)]{dewangan15b}
Dewangan, L.~K., Luna, A., Ojha, D.~K., et al.  2015b, ApJ, 811, 79

\bibitem[Dewangan(2017)]{dewangan17x}
Dewangan, L.~K. 2017, ApJ, 837, 44

\bibitem[Dewangan \& Ojha(2017)]{dewangan17xx}
Dewangan, L.~K., \& Ojha, D.~K. 2017, ApJ, 849, 65

\bibitem[Dewangan et al.(2017a)]{dewangan17}
Dewangan, L.~K., Ojha, D.~K., Zinchenko, I., Janardhan, P., \& Luna, A.  2017a, ApJ, 834, 22

\bibitem[Dewangan et al.(2017b)]{dewangan17b}
Dewangan, L.~K., Ojha, D.~K., \& Zinchenko, I. 2017b, ApJ, 851, 140

\bibitem[Dewangan et al.(2018a)]{dewangan18}
Dewangan, L.~K., Devaraj, R., \& Ojha, D.~K. 2018a, ApJ, 854, 106

\bibitem[Dewangan et al.(2018b)]{dewangan18b}
Dewangan, L.~K., Ojha, D.~K., Zinchenko, I., \& Baug, T. 2018b, ApJ, 861, 19

\bibitem[Elia et al.(2017)]{elia17}
Elia, D., Molinari, S., Schisano, E., et al. 2017, MNRAS, 471, 100

\bibitem[Elmegreen(1998)]{elmegreen98} 
Elmegreen, B.~G., 1998, in ASP Conf. Ser. 148, Origins, ed. C. E. Woodward, J.~M. Shull, \& H.~A. Thronson, Jr. (San Francisco, CA: ASP), 150

\bibitem[Evans et al.(2009)]{evans09}
Evans, N.~J., II, Dunham, M.~M., J\o{}rgensen, J.~K., et al. 2009, ApJS, 181, 321

\bibitem[Flaherty et al.(2007)]{flaherty07}
Flaherty, K.~M., Pipher, J.~L., Megeath, S.~T., et al. 2007, ApJ, 663, 1069

\bibitem[Fujita et al.(2017)]{fujita17}
Fujita, S., Torii, K., Kuno, N., et al. 2017, arXiv:1711.01695

\bibitem[Fukui et al.(2014)]{fukui14} 
Fukui, Y., Ohama, A., Hanaoka, N., et al. 2014, ApJ, 780, 36

\bibitem[Fukui et al.(2016)]{fukui16} 
Fukui, Y., Torii, K., Ohama, A., et al. 2016, ApJ, 820, 26 

\bibitem[Fukui et al.(2018a)]{fukui18a} 
Fukui, Y., Torii, K., Hattori, Y., et al. 2018a, ApJ, 859, 166 

\bibitem[Fukui et al.(2018b)]{fukui18} 
Fukui, Y., Kohno, M., Yokoyama, K., et al. 2018b, PASJ, 70, 41

\bibitem[Furukawa et al.(2009)]{furukawa09} 
Furukawa, N., Dawson, J.~R., Ohama, A., et al. 2009, ApJL, 696, L115 

\bibitem[Getman et al.(2007)]{getman07} 
Getman, K.~V., Feigelson, E.~D., Garmire,G., Broos, P., \& Wang, J. 2007, ApJ, 654, 316 

\bibitem[Gutermuth \& Heyer(2015)]{gutermuth15}
Gutermuth, R.~A., \& Heyer,M. 2015, AJ, 149, 64

\bibitem[Habe \& Ohta(1992)]{habe92} 
Habe, A., \& Ohta, K. 1992, PASJ, 44, 203 

\bibitem[Hartmann et al.(2005)]{hartmann05} 
Hartmann, L., Megeath, S.~T., Allen, L., et al. 2005, ApJ, 629, 881

\bibitem[Haworth et al.(2015a)]{haworth15a} 
Haworth, T.~J., Tasker, E.~J., Fukui, Y., et al. 2015a, MNRAS, 450, 10

\bibitem[Haworth et al.(2015b)]{haworth15b} 
Haworth, T.~J., Shima, K., Tasker, E.~J., et al. 2015b, MNRAS, 454, 1634

\bibitem[Hou et al.(2009)]{hou09} 
\text{Hou, L.~G., Han, J.~L., \& Shi, W.~B. 2009, A\&A 499, 473}

\bibitem[Hou \& Han(2014)]{hou14} 
Hou, L.~G. \& Han, J.~L. 2014, A\&A 569, 125

\bibitem[Hu et al.(2016)]{hu16} 
Hu, B., Menten, K.~M., Wu, Y., et al. 2016, ApJ 833, 18

\bibitem[Inoue \& Fukui(2013)]{inoue13} 
Inoue, T., \& Fukui, Y. 2013, ApJL, 774, 31

\bibitem[Jackson et al.(2006)]{jackson06} 
Jackson, J.~M., Rathborne, J.~M., Shah, R.~Y., et al. 2006, ApJS, 163, 145

\bibitem[Jones \& Dickey(2012)]{jones12} 
Jones, C. \& Dickey, J.~M. 2012, ApJ, 753, 62

\bibitem[Kantharia et al.(2007)]{kantharia07} 
Kantharia, N.~G., Goss, W.~M., Roshi, D.~A., Mohan, N.~R., \& Viallefond, F. 2007, J. Astrophys. Astr., 28, 41

\bibitem[Kendrew et al.(2012)]{kendrew12}
Kendrew S., Simpson R., Bressert E., et al. 2012, ApJ, 755, 71

\bibitem[Kuchar \& Clark(1997)]{kuchar97}
Kuchar, T.~A. \& Clark, F.~O. 1997, ApJ, 488, 224

\bibitem[Lockman(1989)]{lockman89} 
Lockman, F.~J. 1989, ApJS, 71, 469

\bibitem[Mallick et al.(2015)]{mallick15}
Mallick, K.~K., Ojha, D.~K., Tamura, M., et al. 2015, MNRAS, 447, 2307

\bibitem[McClure-Griffiths \& Dickey(2007)]{mcclure07}
McClure-Griffiths, N.~M., \& Dickey, J. 2007, ApJ, 671, 427

\bibitem[Molinari et al.(2010)]{molinari10}
Molinari, S., Swinyard, B., Bally, J., et al., 2010, A\&A, 518, L100

\bibitem[Molinari et al.(2016)]{molinari16}
Molinari, S., Merello, M., Elia, D., Cesaroni, R., Testi, L., \& Robitaille, T. 2016, ApJ, 826L, 8

\bibitem[Motte et al.(2003)]{motte03}
Motte, F., Schilke, P., \& Lis, D.~C. 2003, ApJ, 582, 277

\bibitem[Nguyen Luong et al.(2011)]{nguyen11} 
Nguyen Luong, Q., Motte, F., Schuller, F., et al. 2011, A\&A, 529, A41

\bibitem[Ohama et al.(2010)]{ohama10} 
Ohama, A., Dawson, J.~R., Furukawa, N., et al. 2010, ApJ, 709, 975

\bibitem[Ohama et al.(2018a)]{ohama17b} 
Ohama, A., Kohno, M., Hasegawa, K., et al. 2018a, PASJ, 70, S45

\bibitem[Ohama et al.(2018b)]{ohama17} 
Ohama, A., Kono, M., Fujita, S., et al. 2018b, PASJ, 70, S47

\bibitem[Panagia \& Walmsley(1978)]{panagia78} 
Panagia, N., \& Walmsey, C.~M. 1978, A\&A, 70, 711

\bibitem[Parker et al.(2005)]{parker05}	
Parker, Q.~A., Phillipps, S., Pierce, M. J. et al. 2005, MNRAS, 362, 689

\bibitem[Prinja et al.(1990)]{prinja90}	
Prinja, R.~K., Barlow, M.~J., \& Howarth, I.~D. 1990, ApJ, 361, 607

\bibitem[Rathborne et al.(2011)]{rathborne11}
Rathborne, J.~M., Garay, G., Jackson, J.~M., et al. 2011, ApJ, 741, 120

\bibitem[Reid et al.(2009)]{reid09}
Reid, M.~J., Menten, K.~M., Zheng, X.~W., et al. 2009, ApJ, 700, 137

\bibitem[Reid et al.(2014)]{reid14}
Reid, M.~J., Menten, K.~M., Brunthaler, A., et al. 2014, ApJ, 783, 130

\bibitem[Robitaille et al.(2008)]{robitaille08}
Robitaille T.~P., Meade, M.~R., Babler, B.~L., et al. 2008, AJ, 136, 2413

\bibitem[Sano et al.(2018)]{sano18}
Sano, H., Enokiya, R., Hayashi, K., et al. 2018, PASJ, 70, 43

\bibitem[Sato et al.(2014)]{sato14}
Sato, M., Wu, Y.~W., Immer, K., et al. 2014, ApJ, 793, 72

\bibitem[Schuller et al.(2009)]{schuller09}
Schuller, F., Menten, K.~M., Contreras, Y., et al. 2009, A\&A, 504, 415

\bibitem[Simpson et al.(2012)]{simpson12}
Simpson, R.~J., Povich, M.~S. Kendrew S., et al. 2012, MNRAS, 424, 2442

\bibitem[Stil et al.(2006)]{stil06}
Stil, J.~M., Taylor, A.~R., Dickey, J.~M., et al. 2006, AJ, 132, 1158

\bibitem[Szymczak et al.(2012)]{szymczak12} 	
Szymczak, M., Wolak, P., Bartkiewicz, A., \& Borkowski, K.~M. 2012, AN, 333, 634

\bibitem[Tackenberg et al.(2012)]{tackenberg12}
Tackenberg, J., Beuther, H., Henning, T., et al. 2012, A\&A, 540, 113

\bibitem[Takahira et al.(2014)]{takahira14} 
Takahira, K., Tasker, E.~J., \& Habe, A. 2014, ApJ, 792, 63 

\bibitem[Tan et al.(2014)]{tan14} 	
Tan, J.~C., Beltr\'an, M.~T., Caselli, P., et al. 2014, in Protostars and Planets VI, ed. H. Beuther et al. (Tucson, AZ: Univ. Arizona Press), 149

\bibitem[Thompson et al.(2012)]{thompson12}
Thompson M.~A., Urquhart J.~S., Moore T.~J.~T., Morgan L.~K., 2012, MNRAS, 421, 408

\bibitem[Torii et al.(2017a)]{torii17} 
Torii, K., Hattori, Y., Hasegawa, K., et al. 2017a, ApJ, 835, 142

\bibitem[Torii et al.(2017b)]{torii17x} 
Torii, K., Matsuo, M., Fujita, S., et al. 2017b, PASJ, 70, 51

\bibitem[Takahira et al.(2018)]{takahira18} 
Takahira, K., Shima, K., Habe, A., \& Tasker, E.~J. 2018, PASJ, 70, 58 

\bibitem[Urquhart et al.(2013a)]{urquhart13a} 
Urquhart, J.~S., Moore, T.~J.~T., Schuller, F., et al. 2013a, MNRAS, 431, 1752

\bibitem[Urquhart et al.(2013b)]{urquhart13b} 
Urquhart, J.~S., Thompson, M.~A., Moore, T.~J.~T. et al. 2013b, MNRAS, 435, 400

\bibitem[Urquhart et al.(2018)]{urquhart18} 
Urquhart, J.~S., K\"{o}nig, C., Giannetti, A., et al. 2018, MNRAS, 437, 1059

\bibitem[Walsh et al.(1998)]{walsh98}
Walsh, A.~J., Burton, M.~G., Hyland, A.~R., \& Robinson, G. 1998, MNRAS, 301, 640

\bibitem[Williams et al.(1994)]{williams94} 
Williams, J. P., de Geus, E. J., \& Blitz, L. 1994, ApJ, 428, 693

\bibitem[Xu et al.(2016a)]{xu16}
Xu, Jin-Long, Li, Di, Zhang, Chuan-Peng, et al. 2016a, ApJ, 819,117

\bibitem[Xu et al.(2016b)]{xu16b}
Xu, Y., Reid, M., Dame, T., et al. 2016b, Science Advances, 2, e1600878

\bibitem[Yan et al.(2016)]{yan16}
Yan, Q.~Z., Xu, Y., Zhang, B., et al. 2016, AJ, 152, 117

\bibitem[Zinnecker \& Yorke(2007)]{zinnecker07} 
Zinnecker, H., \& Yorke, H.~W. 2007, ARA\&A, 45, 481 
%
\end{thebibliography}
 \end{document}